\documentclass[preprints,article,accept,moreauthors,pdftex]{Definitions/mdpi} 

\usepackage{caption}
\usepackage{subcaption}
\usepackage{graphicx}
\usepackage{array}
\usepackage{multirow}

\usepackage{titlesec}

\titleformat{\paragraph}
{\normalfont\normalsize}{\theparagraph}{1em}{}
\titlespacing*{\paragraph}
{0pt}{3.25ex plus 1ex minus .2ex}{1.5ex plus .2ex}

\usepackage{tikz}
\usepackage{collcell}
 
\newcommand*{\MinNumber}{0.80}%
\newcommand*{\MidNumber}{0.95} %
\newcommand*{\MaxNumber}{1.00}%

\newcommand{\ApplyGradient}[1]{%
        \ifdim #1 pt > \MidNumber pt
            \pgfmathsetmacro{\PercentColor}{max(min(100.0*(#1 - \MidNumber)/(\MaxNumber-\MidNumber),100.0),0.00)} %
            \hspace{-0.33em}\colorbox{green!\PercentColor!yellow}{#1}
        \else
            \pgfmathsetmacro{\PercentColor}{max(min(100.0*(\MidNumber - #1)/(\MidNumber-\MinNumber),100.0),0.00)} %
            \hspace{-0.33em}\colorbox{red!\PercentColor!yellow}{#1}
        \fi
}

\newcolumntype{R}{>{\collectcell\ApplyGradient}c<{\endcollectcell}}

\firstpage{1} 
\pubvolume{1}
\issuenum{1}
\articlenumber{0}
\pubyear{2022}
\copyrightyear{2022}
\datereceived{} 
\dateaccepted{} 
\datepublished{} 
\hreflink{https://doi.org/} 

\Title{Novel projection schemes for graph-based Light Field coding}

\TitleCitation{Novel projection schemes for graph-based Light Field coding}

\Author{Nguyen Gia Bach $^{1}$\orcidA{}, Chanh Minh Tran $^{1}$\orcidB{}, Tho Nguyen Duc $^{1}$\orcidC{}, Phan Xuan Tan $^{2}$*\orcidD{}, and Eiji Kamioka $^{1}$\orcidE{}}

\AuthorNames{Nguyen Gia Bach, Chanh Minh Tran, Tho Nguyen Duc, Phan Xuan Tan and Eiji Kamioka}

\AuthorCitation{Bach, N.G; Chanh, T.M; Tho, N.D; Tan, P.X; Kamioka, E.}

\address{%
$^{1}$ \quad Graduate School of Engineering and Science, Shibaura Institute of Technology, Tokyo 135-8548, Japan;
nb20502@shibaura-it.ac.jp (C.M.T); nb20501@shibaura-it.ac.jp (T.N.D.)\\
$^{2}$ \quad Department of Information and Communications Engineering, Shibaura Institute of Technology, Tokyo 135-8548, Japan }

\corres{Correspondence: tanpx@shibaura-it.ac.jp (P.X.T.);}

\abstract{In Light Field compression, graph-based coding is powerful to exploit signal redundancy along irregular shapes and obtains good energy compaction. However, apart from high time complexity to process high dimensional graphs, their graph construction method is highly sensitive to the accuracy of disparity information between viewpoints. In real world Light Field or synthetic Light Field generated by computer software, the use of disparity information for super-rays projection might suffer from inaccuracy due to vignetting effect and large disparity between views in the two types of Light Fields respectively. This paper introduces two novel projection schemes resulting in less error in disparity information, in which one projection scheme can also significantly reduce time computation for both encoder and decoder. Experimental results show projection quality of super-pixels across views can be considerably enhanced using the proposals, along with rate-distortion performance when compared against original projection scheme and HEVC-based or JPEG Pleno-based coding approaches.}


\begin{document}

\section{Introduction}

Light Field (LF) is an emerging technology in multimedia research areas that allows capturing different light rays in many directions, emitted from every point of an object or a scene \cite{ref-1}. Hence, it brings significantly improved immersiveness, depth, intensity, color and perspectives from a range of viewpoints. As the result, it reveals promising application opportunities into vast areas such as Virtual Reality (VR), Augmented Reality (AR) \cite{ref-2}, 3D television \cite{ref-3}, biometrics recognition \cite{ref-4}, medical imaging \cite{ref-5}, or post-capture processing techniques like depth estimation and refocusing \cite{ref-6}. However, the rich quality trades off with a high volume of redundant data from both within and between viewpoints, leading to the need of obtaining efficient compression approaches. 

Recently, graph-based coding has proved to be an efficient approach for LF compression \cite{ref-13,ref-14,ref-15} in comparison with conventional 2D image based compression methods, e.g., HEVC \cite{ref-7}, JPEG Pleno \cite{ref-8}. This is because the conventional methods use rectangular blocks which often contain non-uniform intensities or sub regions with different statistical properties. Such non-uniform representation of signal achieves low energy compaction when transformed into frequency domain, leading to higher bitrate required for coding. Meanwhile, graph-based coding can efficiently exploit the redundancy within pixels blocks with irregular shape, adhering closely to object boundaries. More concretely, graphs with arbitrary shapes containing mostly uniform pixel intensities are transformed into frequency domain using Graph Fourier Transform (GFT). As the result, better energy compaction of coefficients can be achieved. Among exiting graph-based LF coding methods, the one in \cite{ref-15} achieves the best rate-distortion performance by proposing graph coarsening and partitioning in a rate-distortion sense. Indeed, in comparison with the methods in \cite{ref-13,ref-14}, this method is capable of reducing graph vertices and obtaining smaller graphs from the original high dimensional graphs. At the same time, it assures that the redundancies within and between views can still be efficiently exploited at some target coding bitrates. This allows the redundancy in bigger pixel regions where the signal is smooth to be efficiently exploited. As the result, high rate-distortion performance can be achieved. However, compared to HEVC Lozenge \cite{ref-7} and JPEG-Pleno \cite{ref-8}, the method in \cite{ref-15} remains outperformed for real LF suffering from vignetting effect and synthetic LF at high bitrates, despite having highest rate-distortion at low bitrates. Additionally, their execution time is reported to be 10 times higher than HEVC for a single LF at the same target quality, mainly due to time complexity of Laplacian eigen-decomposition.

It is believed that the main reason why \cite{ref-15} doesn't perform well on high coding bitrates relates to the error in disparity information used for super-rays projection. To elaborate this point, a closer look on the concept of super-rays as the common support of graph-based LF coding studies is needed. It is an extension to super-pixels over-segmentation in 2D images \cite{ref-24}. In other words, upon views of LF, each super-ray is a group of corresponding super-pixels across all views. The purpose is to group similar light rays coming from the same object in the 3D space to different viewpoints, as an analogy to grouping perceptually similar pixels being close to each other in 2D image. The similarity contains high redundancy, and thus good energy compaction can be obtained in the frequency domain. In details, existing graph-based LF coding studies \cite{ref-13,ref-14,ref-15} segments top-left view into super-pixels, computes the median disparity per super-pixel based on the estimated disparity of top-left view, then applies disparity shift for the projection of a super-pixel from the reference view to remaining views at both encoder and decoder. Due to the similar geometry (structures of objects) and optical characteristics (distance from camera to objects) between the viewpoints, scaling of the disparity value can be used to shift the pixels from one viewpoint to any other viewpoint. This emphasizes the importance of the accuracy of disparity information to the projection of super-pixels.

However, in the case of real LF captured by plenoptic camera or camera array, if the selected reference view suffers from vignetting effect, the estimated disparity would not be accurate, and thus the projection of super-pixels would also suffer errors, leading to incorrect position of corresponding super-pixel in target view. For synthetic LF generated by software, the baseline distance between every two viewpoints has no constraint, and thus it usually has much larger disparity between views compared to real LF, whose baseline is limited by aperture size of a plenoptic camera. Hence, using only one median disparity per super-ray would make the super-ray projection less accurate, particularly when super-pixel size is large.

To this end, in this paper, two novel projection schemes related to selection of reference views for super-ray projection for real LF and synthetic LF, are proposed, to tackle error in disparity information and improve the super-ray projection quality. For real LF with vignetting issue, instead of using top-left view as a reference view, the center view is proposed to be used. This allows the projection to spread out to neighboring views symmetrically in both directions. As the result, the properties of the obtained depth map are preserved. For synthetic LF having large disparity, instead of choosing only a single reference view in top-left corner, multiple views in a sparse distribution is proposed. This allows to perform projection to closer views. As the result, the error of median disparity per super-ray used for projection can be reduced. Moreover, each reference view would be associated with a distinct global graph (a set of all super-ray graphs), and thus the original global graph is divided into smaller sub global graphs. As the result, they can be processed in parallel, decreasing time computation. In order to determine optimal number of views in this proposal, a Lagrangian minimization problem is solved. The purpose is to avoid increasing bitrates during transmission of reference segmentation maps and disparity maps.    

The experimental results demonstrate that by using the proposed projection schemes, higher rate-distortion performance and lower time computation are generally achieved, in comparison with various baselines. The main contributions of the paper are as follows:

\begin{itemize}
\item How vignetting effect results in inaccurate depth estimation, how large disparity between views leads to higher median disparity error for projection, and how these issues affect the projection quality are examined qualitatively and quantitatively.
\item A center view projection scheme is proposed for real LF with large parallax, suffering from vignetting effect in peripheral views, in which the center view is selected as the reference instead of top-left view. This scheme outperforms both original scheme \cite{ref-15} and state-of-the-art coders like HEVC or JPEG Pleno at low and high bitrates.
\item A multiple views projection scheme is proposed for synthetic LF, in which the positions of reference views are optimized by a minimization problem, so that projection quality is improved and inter-views correlations can still be efficiently exploited. In results, this proposal significantly outperforms the original scheme \cite{ref-15} in terms of both Rate Distortion and time computation, by parallel processing sub global graphs with smaller dimensions.
\item A comparative analysis with qualitative and quantitative results is given on rate-distortion performance between the two proposals and original projection scheme \cite{ref-15}, as well as HEVC-Serpentine and JPEG Pleno 4DTM .
\end{itemize}

The rest of the paper is organized as follows: Section \ref{related-work} introduces LF compression categories and recent studies on graph-based LF compression. Section \ref{verify-issues} provides a verification of the issues resulting in the error of disparity information. A detailed description of the two projection schemes is given in Section \ref{proposals}. In Section \ref{results} and \ref{discussion}, experimental results and analysis are discussed to evaluate the performance of proposals. Conclusion is given in Section \ref{conclusion}.

\section{Related Work} \label{related-work}

In this section, the paper first reviews categories of LF representations and their associated compression approach, then further surveys existing studies on graph-based LF compression. The goal is to understand the current progress of LF compression, the potentials of graph-based LF coding and clarify the benefit of graph coarsening and partitioning over other recent graph-based approach, as well as its existing issues.

\subsection{Light Field compression}
LF compression can be generally based on two approaches: compressing the raw lenslet image (2D image) or compressing multiple views (array of 2D images) extracted from the raw data. 

The first category aims at LF with lenslet-based representation, which is a 2D image containing a grid of microlens images, and most of its solutions \cite{ref-9,ref-10,ref-11,ref-12} take advantage of existing HEVC by extending new intra prediction modes exploiting correlation between micro-images, each of which is the captured image from each micro-lens. 

Methods in the second category aim at pseudo-video-sequence based, multiview based, volumetric based, and geometry-assisted based LF representations, all of which can be generated from lenslet acquisition, or multiview acquisition. In the case of raw lenslet acquisition, captured by plenoptic camera, the image is first preprocessed by de-vignetting and demosaicing, then a dense array of views (micro-images or sub-aperture images) are extracted. For multiview acquisition, captured by an array of cameras, multiple views with full parallax can be used directly without preprocessing. The variety of ways the views are stacked together inspire different LF compression approaches.
Pseudo-video-sequence (PVS) based, volumetric based, and multiview based LF representations attempt to stack the views into 1D or 2D array of viewpoints, similar to the concept of traditional 2D or 3D videos, and thus conventional 2D or 3D coders (i.e: HEVC, JP3D \cite{ref-16}, MVC  \cite{ref-17,ref-18,ref-19,ref-20,ref-21}, MVC-HEVC \cite{ref-22}) can be used.

\begin{figure}[H]
\centering
\captionsetup{justification=centering}
\includegraphics[width=10.5 cm]{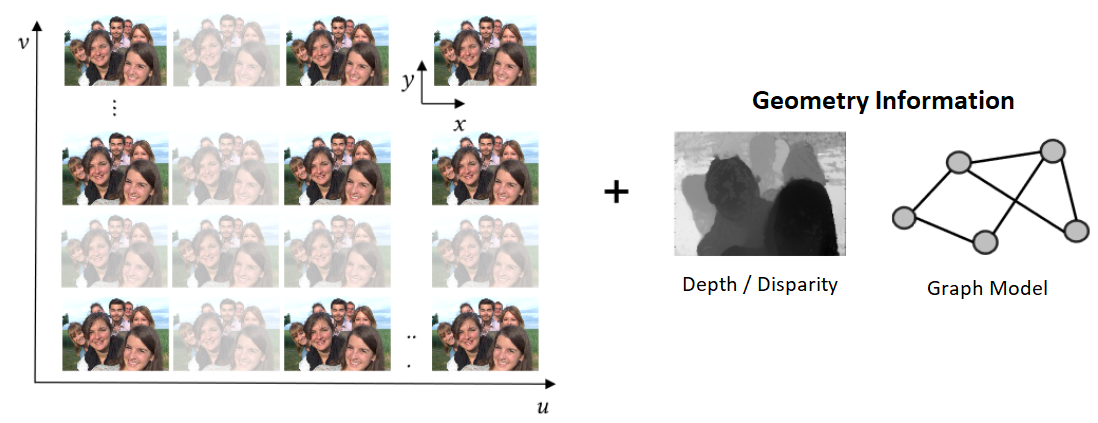}
\caption{Geometry-assisted representation for LF \label{fig1}}
\end{figure}

The most recent geometry-assisted based LF representation does not heavily rely on stacking viewpoints or try to consider the whole LF content as a 2D/3D video, hence it depends less on traditional coders, and has high potential for improvement. Instead, research into this category focuses on key view selection and geometry estimation problems, as depicted in Fig. \ref{fig1}. 

Geometry-assisted based LF representation is accompanied with view synthesis based LF compression, which has been adopted in the 4DPM (4D prediction) mode of JPEG Pleno, a new standard project within the ISO/IEC JTC 1/ SC 29/WG 1 JPEG Committee, specialized in novel image modalities such as textured-plus-depth, light field, point cloud, or holograms. JPEG Pleno implements two strategies to exploit LF redundancy, 4DTM and 4DPM. The 4DTM mode utilizes a 4D transform approach, and targets real LF with high angular view density obtained by plenoptic cameras. Raw LF in lenslet format is first converted into multiview representation, and 4DTM partitions LF into variable-size 4D blocks (two spatial and two angular dimensions), then each block is transformed using 4D DCT. On the other hand, the 4DPM mode divides multiple views of LF into a set of reference views and intermediate views. Texture and geometric depth of reference views are encoded using JPEG 2000, then at the decoder side, a hierarchical depth-based prediction technique is used to obtain depth maps of discarded views, and their textures are warped from the references based on obtained depths. Hence, the 4DPM mode can encode LF very efficiently under reliable depth information. However, at the time of this paper, the 4DPM mode is not yet available in the open source code of JPEG Pleno Reference Software \cite{ref-23}, and thus, the 4DTM mode is used for comparison in this paper instead.

\subsection{Graph-based Light Field coding}
Graph-based Light Field coding falls into the second category of LF compression which compresses a dense array of 2D images (micro-images or sub-aperture images) extracted from the raw lenslet LF, aiming at geometry-assisted based LF representation. Graph vertices are used to describe colors with pixel intensities as graph signals, while graph connections reflect geometry dependencies intra-view or inter-view. The graph signals are transformed into frequency domain to exploit energy compaction using Graph Fourier Transform (GFT), then quantized and encoded to send to the decoder, while the graph support (Laplacian matrix) can be encoded using a separate lossless coder.

\begin{figure}[H]
\begin{minipage}{.5\textwidth}
    \centering
    \captionsetup{justification=centering}
    \includegraphics[width=.8\linewidth]{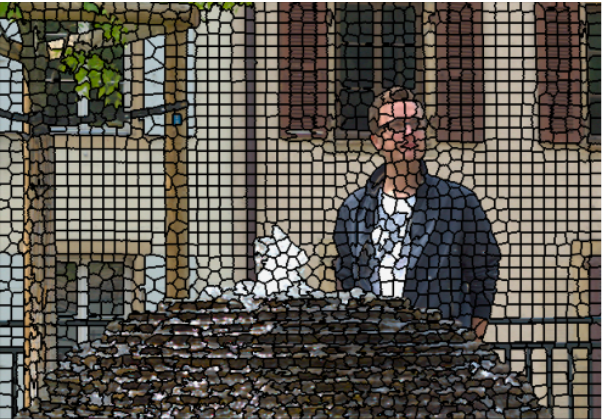}
    \caption{An example of super-pixel segmentation for real world LF on dataset $Fountain\_Vincent\_2$, with number of super-pixels set at 2000 \label{fig4}}
\end{minipage}
\begin{minipage}{.5\textwidth}
    \centering
    \captionsetup{justification=centering}
    \includegraphics[width=.8\linewidth]{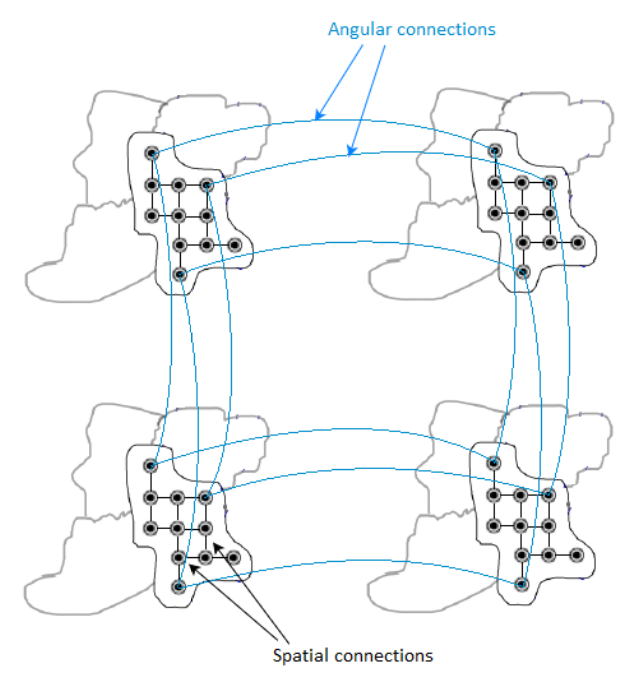}
    \caption{A super-ray graph consisting of spatial graphs connecting pixels within a super-pixel and angular graphs connecting corresponding pixels across views. $I_{1,1}$, $I_{1,2}$, $I_{2,1}$, $I_{2,2}$ are four adjacent views\label{fig5}}
\end{minipage}
\end{figure}

\begin{figure}[H]
\centering
\captionsetup{justification=centering}
\includegraphics[width=.4\linewidth]{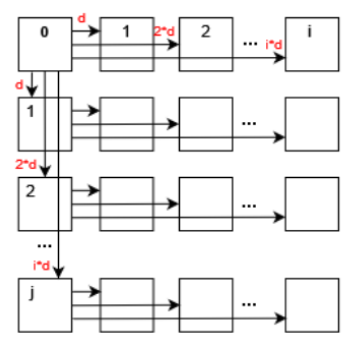}
\caption{Top-left view projection scheme based on median disparity per super-pixel \label{fig6}}
\end{figure}

In \cite{ref-13}, a graph-based solution is proposed with graph support defined on the super-ray segmentation, first introduced in \cite{ref-24} to group light rays of similar color being close in 3D space, as an extension to the concept of super-pixels obtained by SLIC segmentation in 2D image \cite{ref-25}. A super-pixel groups perceptually similar pixels within a view, and a super-ray groups corresponding super-pixels across views, and total super-rays form up the LF image. Their proposal first selects top-left view as the key view, obtains super-pixels segmentation labels using SLIC algorithm \cite{ref-27}, computes its disparity map using method in \cite{ref-28}, then project super-pixel labels to other views based on disparity shift, and construct local graphs of super-rays across all views. Spatial edges connect pixels within a super-pixel, and angular edges form connections between corresponding pixels of the same super-pixel at every four views. An example of the process is illustrated in Fig. \ref{fig4}, Fig. \ref{fig5}, and Fig. \ref{fig6}. Their results have shown to outperform HEVC Lozenge \cite{ref-7} at high bitrates for all real LF datasets, but performs worse at low bitrates. This can be explained by the fact that using limited size for graph support might cope with computational complexity of high dimensional non-separable graph, yet it may not enable GFT to exploit long term spatial or angular redundancy of signal.

In \cite{ref-14}, the authors improve their previous work in \cite{ref-13} by addressing the issue of limited local graph size, and propose sampling and prediction schemes for local graph transform to exploit correlation beyond limits of local graph, without extending graph size. Their proposal first samples the LF data based on graph sampling theory to form a new image of reference samples, then encodes it with conventional 2D image coder with powerful intra-prediction ability. The encoder sends the coded reference image along with only high frequency coefficients of graph transforms. At the decoder side, a prediction mechanism in the pixel domain is introduced to predict the low frequency coefficients using the obtained reference image and high frequency coefficients. Their schemes are designed for quasi-lossless (high quality) coding and have shown substantial RD gain compared to HEVC-Inter Raster scan at lossless mode. However, their performance can drop drastically with lower bitrate, because the prediction scheme is highly dependent on high frequency coefficients, in which a tiny change (i.e. small rounding) may lead to significant reduction in reconstruction quality.

Their most recent study \cite{ref-15} also deals with the high complexity of non-separable graph in \cite{ref-13} using graph coarsening and partitioning, guided by a rate-distortion model for graph optimization. Graph coarsening is performed to reduce the number of vertices inside a super-ray graph, below a threshold leading to acceptable complexity, while retaining basic properties of the graph. If signal approximation of a reduced graph gives too coarse reconstruction of the original signals, or contains two regions with different statistics properties despite having acceptable number of vertices, the local graph is partitioned into two sub-graphs instead. Their experiment results have shown to surpass other state-of-the-art coders like HEVC Lozenge \cite{ref-7} and JPEG Pleno \cite{ref-8} for ideal real LF, but outperformed by most coders at high bitrates (quasi-lossless) for real LF suffered from vignetting effect, and synthetic LF, even though their proposal’s performance still exceed others at low bitrate.

Importantly, both \cite{ref-14} and \cite{ref-15} implement the same super-ray projection mechanism as in \cite{ref-13}. They select the top-left view as the reference view, compute its disparity map and segmentation map, then project the super-pixels labels to all other views based on the median disparity per super-pixel, as illustrated in Fig. \ref{fig6}. The projection scheme proceeds row by row, with horizontal projection from left to right in each row, and one vertical projection from above for the first view of every row. However, top-left view might not be the optimal selection for reference view on real LF due to vignetting, and choosing one median disparity per super-ray might incur high disparity error on synthetic LF, if every pixel in super-pixel has high disparity, especially when the super-pixel is large. This research's purpose is to clarify how the above issues bring negative impact on disparity information, then to propose two novel projection schemes to mitigate the issues, and obtain enhanced projection quality of super-pixels. The main focus is to improve the most recent graph-based solution \cite{ref-15} with already better performance among the other two approaches \cite{ref-13, ref-14}, so that it can perform well on all range of coding bitrates for all types of LF, using the two proposed projection schemes.

\section{Impact of disparity information on projection quality} \label{verify-issues}

In this section, the issues related to affecting reconstructed view quality in \cite{ref-15} are shown as follows, 
\begin{itemize}
\item How vignetted real LF affects its disparity estimation ?
\item How synthetic LF with large disparity leads to higher median disparity error ?
\item How both issues affect the quality of super-ray projection ? To support the verification, SSIM metric \cite{ref-36} was used to compute the projection quality for each view with top-left view projection.
\end{itemize}

\subsection{Vignetting effect degrades disparity estimation}
Vignetting has been extensively surveyed regarding its impact on traditional 2D stereo correspondence problems \cite{ref-26} among other radiometric differences such as image noises, different camera settings, etc. Stereo correspondence or stereo matching is an active topic in computer vision with the goal to estimate depth information for 2D images from an image pair. However, optical flow based approach has shown to outperform stereo based algorithms \cite{ref-29, ref-30} in the task of depth estimation for LF in recent literature \cite{ref-27, ref-28}. This section provides a demonstration on how vignetting impacts optical flow based LF depth estimation in a subjective manner.

\begin{figure}[H]
\centering
\begin{subfigure}{.33\textwidth}
    \centering
    \captionsetup{justification=centering}
    \includegraphics[width=.7\linewidth]{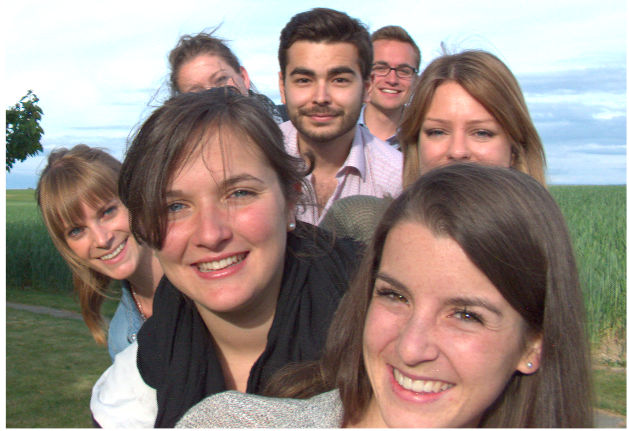}
    \caption{Top-left view \label{fig7a}}
\end{subfigure}%
\begin{subfigure}{.33\textwidth}
    \centering
    \captionsetup{justification=centering}
    \includegraphics[width=.7\linewidth]{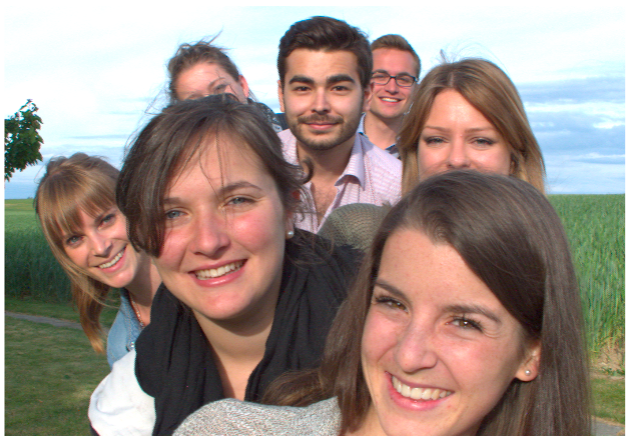}
    \caption{Center view \label{fig7b}}
\end{subfigure}
\captionsetup{justification=centering}
\caption{Samples of views in dataset $Friends$}
\label{fig7}
\end{figure}

\begin{figure}[H]
\centering
\begin{subfigure}{.33\textwidth}
    \centering
    \captionsetup{justification=centering}
    \includegraphics[width=.7\linewidth]{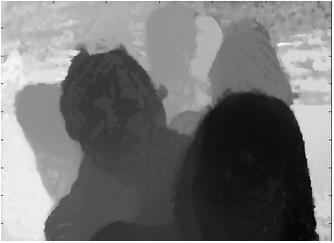}
    \caption{Estimated disparity map of top-left view using \cite{ref-28} \label{fig8a}}
\end{subfigure}%
\begin{subfigure}{.33\textwidth}
    \centering
    \captionsetup{justification=centering}
    \includegraphics[width=.7\linewidth]{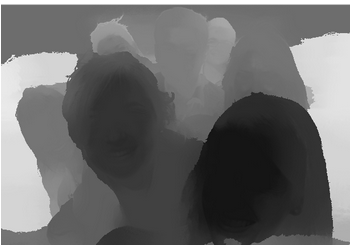}
    \caption{Estimated disparity map of center view using \cite{ref-29} \label{fig8b}}
\end{subfigure}%
\begin{subfigure}{.33\textwidth}
    \centering
    \captionsetup{justification=centering}
    \includegraphics[width=.7\linewidth]{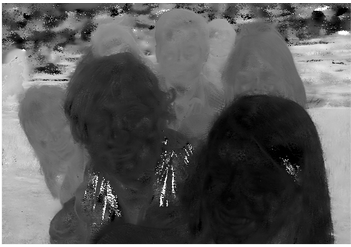}
    \caption{Estimated disparity map of center view using \cite{ref-30} \label{fig8b}}
\end{subfigure}
\captionsetup{justification=centering}
\caption{Disparity maps obtained using optical-flow based and stereo-based depth estimators in dataset $Friends$}
\label{fig8}
\end{figure}

Due to inability to efficiently capture light rays in peripheral lens, plenoptic cameras usually produce vignetted border views in a multiview based representation of a LF content captured at wide angle, which is essential to provide high parallax. First, considering the case of real LF with medium parallax (9x9 views), this experiment examines the difference between the top-left and center views of dataset $Friends$ \cite{ref-32} qualitatively. As depicted in Fig. \ref{fig7}, the two images are almost identical, with a tiny shift in position of objects, but no visual distortion in terms of colors or blurring occurs. The depth map of top-left view is then computed using optical flow based method in \cite{ref-28} and compared with their original result, also with two other state-of-the-art stereo based disparity estimators, \cite{ref-29} and \cite{ref-30}. From Fig. \ref{fig8}, it’s apparent that the obtained disparity map adheres closely to the basic depth properties which all methods share in common.

\begin{figure}[H]
\centering
\begin{subfigure}{.33\textwidth}
    \centering
    \captionsetup{justification=centering}
    \includegraphics[width=.7\linewidth]{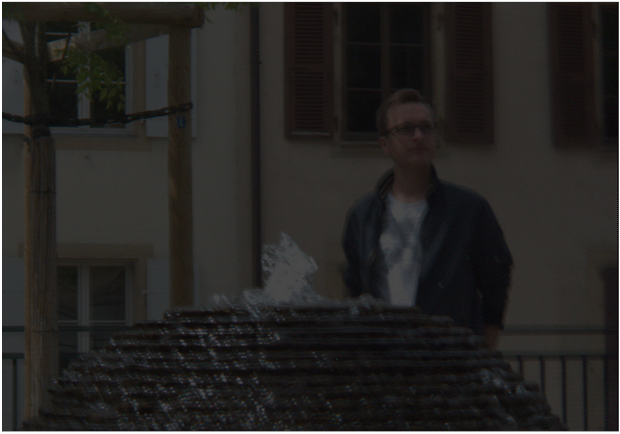}
    \caption{Top-left view (original) \label{fig9a}}
\end{subfigure}%
\begin{subfigure}{.33\textwidth}
    \centering
    \captionsetup{justification=centering}
    \includegraphics[width=.7\linewidth]{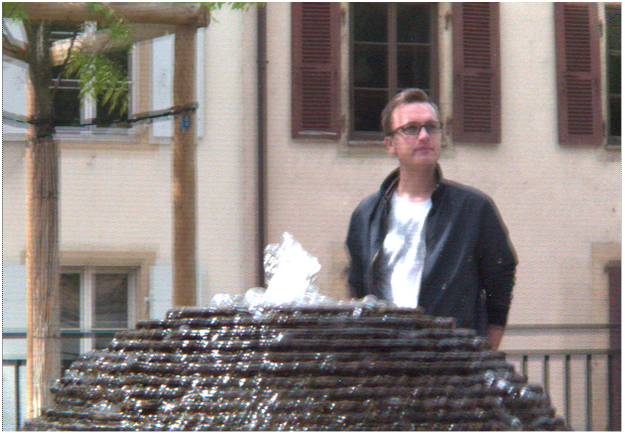}
    \caption{Top-left view (de-vignetted using gamma correction) \label{fig9b}}
\end{subfigure}%
\begin{subfigure}{.33\textwidth}
    \centering
    \captionsetup{justification=centering}
    \includegraphics[width=.7\linewidth]{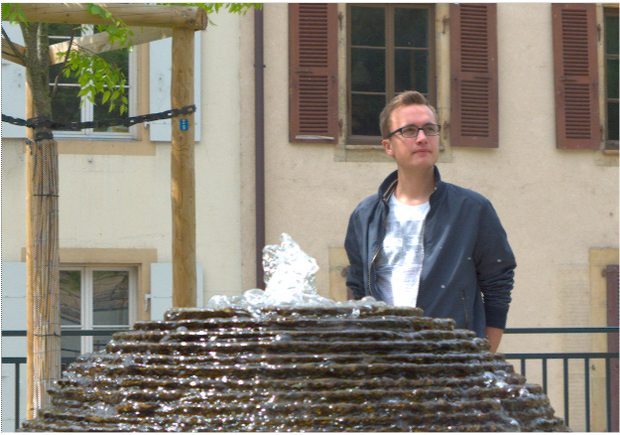}
    \caption{Center view \label{fig9c}}
\end{subfigure}
\captionsetup{justification=centering}
\caption{Samples of views in dataset $Fountain\_Vincent\_2$}
\label{fig9}
\end{figure}

\begin{figure}[H]
\centering
\begin{subfigure}{.24\textwidth}
    \centering
    \captionsetup{justification=centering}
    \includegraphics[width=.9\linewidth]{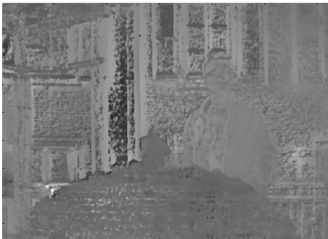}
    \caption{Estimated disparity map of top-left view (gamma corrected) using \cite{ref-28} \label{fig10a}}
\end{subfigure}%
\begin{subfigure}{.24\textwidth}
    \centering
    \captionsetup{justification=centering}
    \includegraphics[width=.9\linewidth]{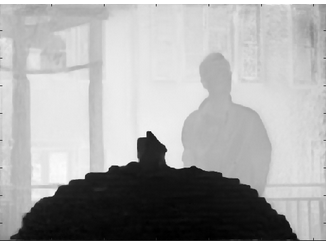}
    \caption{Estimated disparity map of center view using \cite{ref-28} \label{fig10a}}
\end{subfigure}%
\begin{subfigure}{.24\textwidth}
    \centering
    \captionsetup{justification=centering}
    \includegraphics[width=.9\linewidth]{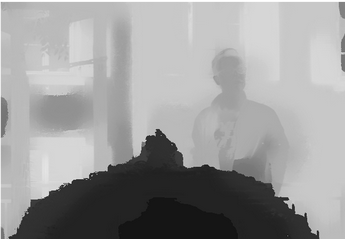}
    \caption{Estimated disparity map of center view using \cite{ref-29} \label{fig10b}}
\end{subfigure}%
\begin{subfigure}{.24\textwidth}
    \centering
    \captionsetup{justification=centering}
    \includegraphics[width=.9\linewidth]{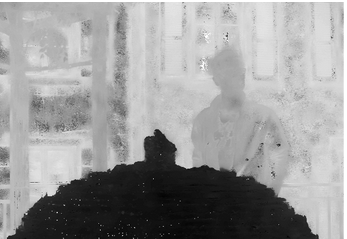}
    \caption{Estimated disparity map of center view using \cite{ref-30} \label{fig10b}}
\end{subfigure}
\captionsetup{justification=centering}
\caption{Disparity maps obtained using optical-flow based and stereo-based depth estimators in dataset $Fountain\_Vincent\_2$}
\label{fig10}
\end{figure}

However, in the case of real LF with high parallax (13x13 views), the obtained depth map might lose these properties due to vignetting effect, even after de-vignetted by color correction methods. A comparison between the top-left view and center view of dataset $Fountain\_Vincent\_2$ \cite{ref-32} with 13x13 views is shown in Fig. \ref{fig9a} and Fig. \ref{fig9c}. The original top-left view can be seen with significant degradation in intensity and color. Gamma correction can be used for color calibration, and thus help reduce vignetting considerably, as seen from Fig. \ref{fig9b}. However, a close look at the calibrated top-left view reveals object details are blurry with minor distortion in the colors, yet its estimated disparity map is severely affected. It is much worse compared to the output in center view using the same optical flow based \cite{ref-28} or stereo based \cite{ref-29,ref-30} depth estimation, illustrated in Fig. \ref{fig10}. For example, the fountain, which is closest to the camera, now shares the same depth as some of objects in the background (i.e. tree), whereas a segment of the wall now becomes the closest, according to its depth. The results have shown even a minor vignetting effect can result in significant deterioration of the estimated depth map in peripheral views.

\subsection{Synthetic LF with large disparity leads to high median disparity error for a super-pixel}

While real LF captured using plenoptic camera has baseline limited by the aperture size of the camera lens, synthetic LF can be generated using graphics software without baseline restraint, enabling wide parallax for the viewer. Therefore, synthetic LF may have much larger disparity between views than  real LF, leading to higher median disparity error per super-ray. This can considerably affect the projection of super-pixel locations from reference view to remaining views. In Table \ref{tbl1}, the previews of two real LF and two synthetic LF are displayed, along with their optical flow between a pair of views, and the disparity range. The intensity of color in the optical flow is proportional to the disparity. And the disparity range displays maximum and minimum of disparity value in the whole LF. It can be seen that synthetic LF has distinctively more intense optical flow and wider range of disparity.

\begin{table}[h!]
  \centering
  \begin{tabular}{ | m{3.5cm} | m{3cm} | m{3cm} | m{2.5cm} | }
    \hline
    Datasets & Center view & Optical Flow & Disparity range\\ \hline
    $Fountain\_Vincent\_2$ \cite{ref-32} (real LF)
    &
    \begin{minipage}{.2\textwidth}
      \includegraphics[width=2cm]{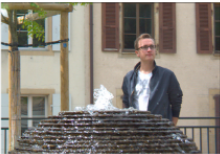}
    \end{minipage}
    & 
    \begin{minipage}{.2\textwidth}
      \includegraphics[width=2cm]{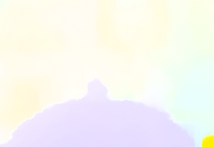}
    \end{minipage}
    & 
    $-0.495 \rightarrow 0.798$
    \\ \hline
    $Danger\_de\_Mort$ \cite{ref-32} (real LF)
    &
    \begin{minipage}{.2\textwidth}
      \includegraphics[width=2cm]{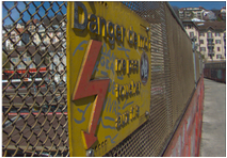}
    \end{minipage}
    & 
    \begin{minipage}{.2\textwidth}
      \includegraphics[width=2cm]{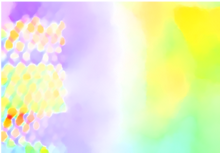}
    \end{minipage}
    & 
    $-1.306 \rightarrow 0.683$
    \\ \hline
    $Greek$ \cite{ref-33} (synthetic LF)
    &
    \begin{minipage}{.2\textwidth}
      \includegraphics[width=2cm]{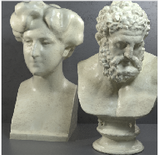}
    \end{minipage}
    & 
    \begin{minipage}{.2\textwidth}
      \includegraphics[width=2cm]{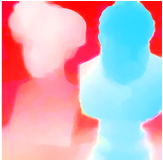}
    \end{minipage}
    & 
    $-2.880 \rightarrow 3.637$
    \\ \hline
    $Sideboard$ \cite{ref-33} (synthetic LF)
    &
    \begin{minipage}{.2\textwidth}
      \includegraphics[width=2cm]{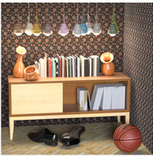}
    \end{minipage}
    & 
    \begin{minipage}{.2\textwidth}
      \includegraphics[width=2cm]{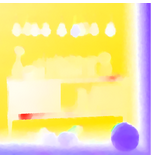}
    \end{minipage}
    & 
    $-1.513 \rightarrow 1.845$
    \\ \hline
  \end{tabular}
  \captionsetup{justification=centering}
  \caption{Comparison of disparity range between real LF and synthetic LF}\label{tbl1}
\end{table}

\begin{figure}[H]
\centering
\captionsetup{justification=centering}
\includegraphics[width=.4\linewidth]{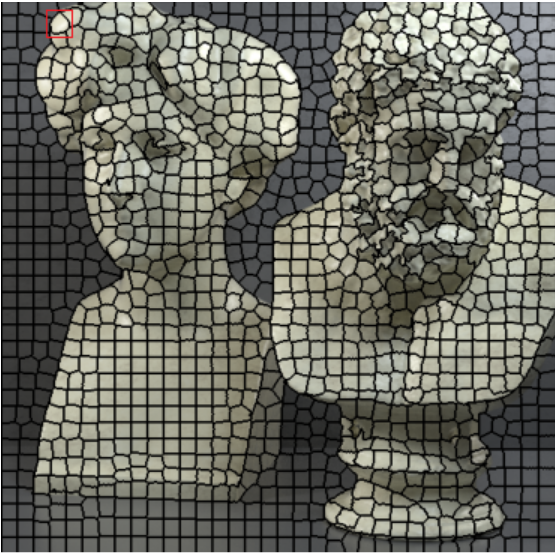}
\caption{An example of super-pixel segmentation for synthetic LF on dataset Greek, with the number of super-pixels set at 1200. The selection of an arbitrary super-pixel is highlighted in red. \label{fig11}}
\end{figure}

In order to verify this issue, a mathematical explanation is given. This scenario considers the projection scheme based on disparity shift, which was used by the authors in \cite{ref-13, ref-14, ref-15}. As illustrated in Fig. \ref{fig6}, disparity shift scheme increases disparity proportionally to the distance between the target and reference view. Our goal is to verify that the further the target view, the higher the median disparity error, hence worse projection of super-pixel labels. Let’s consider an arbitrary super-pixel in the reference view (top-left), depicted in Fig. \ref{fig11}. 

Denote all disparity values of every pixel in this super-pixel as $D={d_1,d_2, ..., d_n}$ where n is the total number of pixels, and $d_m$ is the median disparity value of this set. Since a super-pixel follows well to any object’s boundary, it’s safe to assume that no super-pixel contains pixels of two separate objects at two distinct depth planes, and thus the disparity variation is smooth, or a normal distribution of disparity values in a super-pixel is obtained most often. Therefore, without loss of generality for all super-pixels, this verification considers the median disparity is equal to the mean value, as equation (\ref{eq:1}),

\begin{equation} \label{eq:1}
d_m = \frac{\sum_{i=1}^{n}d_i}{n}
\end{equation}

The median disparity error is calculated as follows,

\begin{equation*} \label{eq:2}
mse = \frac{1}{n}\sum_{i=1}^{n}(d_i-d_m)^2 = \frac{1}{n}\sum_{i=1}^{n}\triangle^2
\end{equation*}

Denote $mse_1$, $\triangle_1$ as the median disparity error and the disparity error, respectively calculated at view 1 (reference view), and $mse_k$, $\triangle_k$ at view k. Since disparity of a pixel at view k is k times higher than at view 1, the following equations are derived,

\begin{equation*} \label{eq:3}
mse_1 = \frac{1}{n}\sum_{i=1}^{n}\triangle_1^2
\end{equation*}

\begin{equation*} \label{eq:4}
mse_k = \frac{1}{n}\sum_{i=1}^{n}\triangle_k^2
\end{equation*}

\begin{equation*} \label{eq:5}
\triangle_1 = d_i-d_m = d_i-\frac{\sum_{i=1}^{n}d_i}{n}
\end{equation*}

\begin{equation*} \label{eq:6}
\triangle_k = k*d_i-d'_m = k*d_i-\frac{\sum_{i=1}^{n}k*d_i}{n} = k*(d_i-\frac{\sum_{i=1}^{n}d_i}{n}) = k*\triangle_1
\end{equation*}

\begin{equation} \label{eq:7}
\Rightarrow mse_k = \frac{1}{n}\sum_{i=1}^{n}k^2*\triangle_1^2 = k^2*\frac{1}{n}\sum_{i=1}^{n}\triangle_1^2 = k^2*mse_1
\end{equation}

Hence, if using the original projection scheme proposed by the authors \cite{ref-13, ref-14, ref-15}, the median disparity error of a specific super-pixel at target view $k$ is $k^2$ higher than the reference view. As the consequences, the further the target view, the more inaccurate that super-pixel is relocated. Furthermore, the issue becomes worse when the disparity is already large, as in synthetic LF.

\subsection{Inaccurate disparity information leads to poor super-ray projection}

To evaluate how super-ray projection scheme used in \cite{ref-13, ref-14, ref-15} are affected by disparity error in a quantitative manner, SSIM \cite{ref-36} metric was used to measure the similarity of the segmentation labels in a target view, between ground truth labels (segmented using SLIC algorithm \cite{ref-27}) and projected labels from reference view, with projection scheme depicted in Fig. \ref{fig6}.
Specifically, the evaluation formula is given in equation (\ref{eq:8}),

\begin{equation} \label{eq:8}
quality = SSIM(L_{i,j}, L'_{i,j})
\end{equation}

where $i$, or $j$ is the location of the target view in 2D array of views; $L_{i,j}$ is denoted as “ground truth” image where every pixel intensity is the label value of the super-pixel it belongs to, assigned by SLIC; $L'_{i,j}$ is also an image of labels, but assigned by super-ray projection from reference view. SSIM value range is between 0 and 1, with 1 meaning best quality, or the two images are identical, and the lower the worse.

\subsubsection{For real world LF with high parallax (vignetting)}

First, the real world LF dataset $Fountain\_Vincent\_2$ \cite{ref-32} 13x13 views was considered. This dataset suffers from vignetting, which then leads to inaccurate disparity map. The ground truth labels of each view were obtained using python SLIC segmentation library, with parameter values as $compactness=30$ and $n\_segments=2000$. The projection quality of 13x13 views was computed, as shown in Table \ref{tbl2}. Results have shown that the view with highest quality is apparently the reference view (top-left), since no projection was made, and the quality gradually deteriorates when the projection was made to further target views. The disparity error accumulated for lower views being selected as references for horizontal projection in the first column, and the views at bottom right corner obtained the worst quality, which means the reconstructed super-pixels of those views might have been located at highly inaccurate positions, hence the distortion. Similarly, the projection quality of dataset $Danger\_de\_Mort$ \cite{ref-32} is given in Table \ref{tbl3}.

\begin{table}[H]
\centering
\begin{tabular}{*{13}{R}}
 \hline
 1.000 & 0.971 & 0.963 & 0.958 & 0.953 & 0.952 & 0.950 & 0.950 & 0.950 & 0.945 & 0.945 & 0.942 & 0.941 \\ \hline
0.956 & 0.961 & 0.955 & 0.951 & 0.947 & 0.947 & 0.945 & 0.946 & 0.945 & 0.942 & 0.941 & 0.939 & 0.936 \\ \hline
0.943 & 0.941 & 0.939 & 0.939 & 0.935 & 0.934 & 0.932 & 0.932 & 0.931 & 0.932 & 0.930 & 0.928 & 0.926 \\ \hline
0.935 & 0.933 & 0.930 & 0.930 & 0.929 & 0.927 & 0.927 & 0.923 & 0.923 & 0.925 & 0.923 & 0.922 & 0.921 \\ \hline
0.921 & 0.920 & 0.922 & 0.923 & 0.921 & 0.920 & 0.918 & 0.917 & 0.915 & 0.916 & 0.915 & 0.913 & 0.915 \\ \hline
0.914 & 0.913 & 0.915 & 0.915 & 0.915 & 0.914 & 0.910 & 0.909 & 0.909 & 0.909 & 0.908 & 0.909 & 0.909 \\ \hline
0.907 & 0.907 & 0.905 & 0.906 & 0.906 & 0.905 & 0.903 & 0.903 & 0.904 & 0.903 & 0.903 & 0.903 & 0.902 \\ \hline
0.902 & 0.902 & 0.900 & 0.900 & 0.900 & 0.899 & 0.899 & 0.899 & 0.899 & 0.899 & 0.897 & 0.897 & 0.896 \\ \hline
0.895 & 0.897 & 0.895 & 0.896 & 0.896 & 0.896 & 0.896 & 0.895 & 0.895 & 0.894 & 0.893 & 0.892 & 0.891 \\ \hline
0.892 & 0.893 & 0.892 & 0.890 & 0.891 & 0.891 & 0.892 & 0.892 & 0.890 & 0.890 & 0.889 & 0.888 & 0.889 \\ \hline
0.886 & 0.888 & 0.887 & 0.884 & 0.885 & 0.885 & 0.886 & 0.886 & 0.886 & 0.886 & 0.884 & 0.882 & 0.883 \\ \hline
0.877 & 0.883 & 0.885 & 0.882 & 0.883 & 0.882 & 0.882 & 0.881 & 0.882 & 0.881 & 0.880 & 0.880 & 0.880 \\ \hline
0.873 & 0.875 & 0.876 & 0.880 & 0.880 & 0.878 & 0.879 & 0.878 & 0.877 & 0.876 & 0.877 & 0.877 & 0.876 \\
 \hline
\end{tabular}
\captionsetup{justification=centering}
\caption{Projection quality of 13x13 views in dataset $Fountain\_Vincent\_2$ using top-left view projection scheme. Reference view with absolute quality is highlighted in green.}\label{tbl2}
\end{table}

\begin{table}[H]
\centering
\begin{tabular}{*{13}{R}} 
 \hline
1.000 & 0.969 & 0.961 & 0.959 & 0.951 & 0.951 & 0.948 & 0.952 & 0.949 & 0.943 & 0.943 & 0.941 & 0.939 \\ \hline
0.955 & 0.959 & 0.953 & 0.949 & 0.946 & 0.945 & 0.944 & 0.945 & 0.943 & 0.941 & 0.939 & 0.937 & 0.934 \\ \hline
0.941 & 0.940 & 0.938 & 0.938 & 0.933 & 0.933 & 0.930 & 0.931 & 0.929 & 0.930 & 0.929 & 0.926 & 0.925 \\ \hline
0.933 & 0.932 & 0.928 & 0.929 & 0.929 & 0.926 & 0.925 & 0.921 & 0.922 & 0.923 & 0.921 & 0.921 & 0.920 \\ \hline
0.920 & 0.918 & 0.920 & 0.921 & 0.919 & 0.919 & 0.917 & 0.915 & 0.913 & 0.915 & 0.913 & 0.912 & 0.913 \\ \hline
0.913 & 0.914 & 0.914 & 0.914 & 0.913 & 0.912 & 0.911 & 0.908 & 0.907 & 0.908 & 0.909 & 0.908 & 0.907 \\ \hline
0.906 & 0.905 & 0.904 & 0.904 & 0.905 & 0.903 & 0.901 & 0.902 & 0.903 & 0.901 & 0.901 & 0.901 & 0.900 \\ \hline
0.900 & 0.901 & 0.898 & 0.901 & 0.899 & 0.897 & 0.898 & 0.897 & 0.897 & 0.897 & 0.897 & 0.894 & 0.893 \\ \hline
0.893 & 0.896 & 0.894 & 0.895 & 0.895 & 0.894 & 0.894 & 0.894 & 0.893 & 0.891 & 0.892 & 0.889 & 0.888 \\ \hline
0.891 & 0.891 & 0.892 & 0.889 & 0.890 & 0.890 & 0.890 & 0.891 & 0.888 & 0.889 & 0.888 & 0.886 & 0.886 \\ \hline
0.888 & 0.887 & 0.886 & 0.882 & 0.884 & 0.883 & 0.885 & 0.883 & 0.884 & 0.883 & 0.883 & 0.879 & 0.882 \\ \hline
0.876 & 0.881 & 0.884 & 0.880 & 0.882 & 0.880 & 0.881 & 0.880 & 0.879 & 0.878 & 0.879 & 0.878 & 0.877 \\ \hline
0.872 & 0.874 & 0.875 & 0.878 & 0.879 & 0.877 & 0.877 & 0.877 & 0.874 & 0.873 & 0.876 & 0.874 & 0.873 \\ \hline
\end{tabular}
\captionsetup{justification=centering}
\caption{Projection quality of 13x13 views in dataset $Danger\_de\_Mort$ using top-left view projection scheme. Reference view with absolute quality is highlighted in green.}\label{tbl3}
\end{table}

\subsubsection{For synthetic LF with high median disparity error per super-ray}

The projection quality was examined in synthetic dataset $Greek$ \cite{ref-33} 9x9 views using SLIC with the same parameters as for real LF datasets. Results in Table \ref{tbl4} reveal that the quality of each view deteriorated much faster in both directions, due to large disparity between the views. Despite having fewer views than vignetted real LF, dataset Greek ended up having worse projection quality at bottom right views. Similarly, Table \ref{tbl5} illustrates projection quality in synthetic dataset $Sideboard$ \cite{ref-33}.

\begin{table}[H]
\centering
\begin{tabular}{*{9}{R}} 
 \hline
1.000 & 0.958 & 0.941 & 0.931 & 0.927 & 0.924 & 0.921 & 0.922 & 0.918 \\ \hline
0.941 & 0.932 & 0.925 & 0.919 & 0.916 & 0.913 & 0.911 & 0.910 & 0.906 \\ \hline
0.915 & 0.912 & 0.908 & 0.906 & 0.904 & 0.901 & 0.900 & 0.898 & 0.896 \\ \hline
0.898 & 0.897 & 0.898 & 0.896 & 0.895 & 0.893 & 0.891 & 0.889 & 0.887 \\ \hline
0.889 & 0.889 & 0.887 & 0.886 & 0.886 & 0.885 & 0.883 & 0.882 & 0.883 \\ \hline
0.887 & 0.886 & 0.884 & 0.884 & 0.882 & 0.882 & 0.882 & 0.882 & 0.883 \\ \hline
0.889 & 0.888 & 0.886 & 0.886 & 0.884 & 0.884 & 0.883 & 0.884 & 0.884 \\ \hline
0.884 & 0.880 & 0.880 & 0.878 & 0.878 & 0.878 & 0.879 & 0.879 & 0.875 \\ \hline
0.877 & 0.875 & 0.872 & 0.872 & 0.871 & 0.871 & 0.872 & 0.870 & 0.870 \\
 \hline
\end{tabular}
\captionsetup{justification=centering}
\caption{Projection quality of 9x9 views in dataset $Greek$ using single-view projection scheme. Reference view with absolute quality is highlighted in green.}\label{tbl4}
\end{table}

\begin{table}[H]
\centering
\begin{tabular}{*{13}{R}} 
 \hline
1.000 & 0.952 & 0.927 & 0.914 & 0.911 & 0.906 & 0.901 & 0.901 & 0.895 \\
 \hline
0.949 & 0.934 & 0.921 & 0.911 & 0.906 & 0.902 & 0.899 & 0.897 & 0.892 \\
 \hline
0.919 & 0.914 & 0.907 & 0.898 & 0.892 & 0.891 & 0.886 & 0.884 & 0.885 \\
 \hline
0.904 & 0.897 & 0.891 & 0.888 & 0.882 & 0.884 & 0.881 & 0.877 & 0.877 \\
 \hline
0.882 & 0.885 & 0.881 & 0.878 & 0.876 & 0.877 & 0.874 & 0.874 & 0.868 \\
 \hline
0.877 & 0.875 & 0.873 & 0.872 & 0.870 & 0.870 & 0.869 & 0.866 & 0.863 \\
 \hline
0.867 & 0.869 & 0.869 & 0.868 & 0.867 & 0.867 & 0.865 & 0.862 & 0.863 \\
 \hline
0.863 & 0.862 & 0.863 & 0.864 & 0.863 & 0.865 & 0.864 & 0.861 & 0.859 \\
 \hline
0.860 & 0.860 & 0.860 & 0.859 & 0.861 & 0.862 & 0.860 & 0.858 & 0.857 \\
 \hline
\end{tabular}
\captionsetup{justification=centering}
\caption{Projection quality of 9x9 views in dataset $Sideboard$ using single-view projection scheme. Reference view with absolute quality is highlighted in green.}\label{tbl5}
\end{table}

\section{Proposals} \label{proposals}

In this paper, two novel projection schemes are proposed for real LF and synthetic LF as follows: 
\begin{itemize}
    \item For real LF with many viewpoints suffering from vignetting effect, the proposed approach is that super-ray projection is carried out on the center view as the reference, then spreads out to surrounding views, instead of top-left one with inaccurate disparity. 
    \item For synthetic LF with large disparity, a projection scheme using multiple views in a sparse distribution as references is proposed, aiming to reduce the distance between target and reference views. In addition, using multiple reference views can create multiple sub global graphs which are processed simultaneously. This allows to mitigate computational time for both encoder and decoder. 
\end{itemize}

\subsection{Center-view projection scheme}
\begin{figure}[H]
\centering
\captionsetup{justification=centering}
\includegraphics[width=.3\linewidth]{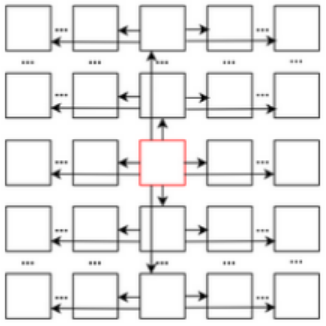}
\caption{Center-view projection scheme\label{fig12}}
\end{figure}

The proposed center-view projection scheme is illustrated in Fig. \ref{fig12}. The purpose of this proposal is to improve rate distortion performance of [15] in real world LF data with large parallax, which is suffered from vignetting in peripheral views. From a center view, the projection spreads out to neighboring views, instead of proceeding row by row in one direction, as in \cite{ref-13, ref-14, ref-15}. Specifically, for a $NxN$ views real LF, this scheme performs a horizontal projection from the center view $I_{\frac{N+1}{2},\frac{N+1}{2}}$ to remaining $\frac{N}{2}$ views symmetrically on the center row $R_\frac{N+1}{2}$. Vertical projection is also carried out in the center column symmetrically in both directions, with  $I_{\frac{N+1}{2},\frac{N+1}{2}}$ as reference. Then for each remaining $N-1$ rows, its center view is now used as reference for the horizontal projection, covering all views of remaining  $N-1$ columns. This projection scheme not only avoids inaccurate disparity estimation in top-left view due to vignetting, but also allows projection to closer views (half the distance compared to top-left view projection), hence quality of more views can be improved.

\subsection{Multiple views projection scheme  }
The purpose of this proposal is to improve rate distortion performance and reduce time computation when applying the approach in \cite{ref-15} into synthetic LF data with large disparity between views. Equation (\ref{eq:7}) has shown that the median disparity error of any super-pixel in a target view is $k^2$ higher than of corresponding super-pixel in reference view, in which target view is the $k^{th}$ view away from the reference. This negatively affects the projection quality. In addition, using a single view as a reference will result in a single global graph with very high dimension, which leads to high encoding and decoding time. Hence, to deal with these two problems at the same time, a novel idea is to increase the number of reference views in a row or column. Thereby, the distance from reference views to the to-be-projected views will be decreased. At the same time, multiple smaller global graphs can be created, which enable leveraging the power of parallel computing to improve the execution time.

The question is how many references should be sufficient. Too many references would lead to inability to efficiently exploit angular correlation across views of the whole LF, whereas few references would cause each reference view to project to further views, and thus increase projection error. Another interpretation of this question is how many views should each reference view project to, in a row or a column.

The question can be answered by finding a target view with worst projection quality, while being close to the reference view as much as possible, then use its ground truth segmented labels as a new reference view for a new projection chain. This can be interpreted as a Lagrangian minimization problem,

\begin{equation} \label{eq:9}
min (k[x] + \lambda*SSIM[x])
\end{equation}
or
\begin{equation} \label{eq:10}
min (SSIM[x] + \lambda*k[x])
\end{equation}

, where $k$ can be considered as an array of distances between target and reference view in a single direction (with number of views as unit) $k={k_1,k_2, ..., k_{n-1}}$; $SSIM$ is an array of projection quality computed using equation \ref{eq:8} in the same direction (multiplied with 100 to be on the same scale as k) $SSIM={q_1,q_2, ..., q_{n-1}}$; and n is the number of views for that row or column. k and SSIM arrays exclude the reference view. Equation (\ref{eq:9}) or (\ref{eq:10}) can be re-expressed as,

\begin{equation} \label{eq:11}
min_x \; k[x] \quad w.r.t \quad SSIM[x] = SSIM_{target} = const
\end{equation}
or
\begin{equation} \label{eq:12}
min_x \; SSIM[x] \quad w.r.t \quad k[x] = k_{target} = const
\end{equation}

The optimal Lagrange multiplier $\lambda^*$ is unknown in advance, and can be varied with the desired distance of views ($k_{target}$) or quality ($SSIM_{target}$).

The optimal solutions to equation (\ref{eq:11}) or (\ref{eq:12}) are the optimal points ($k[x^*]$, $SSIM[x^*]$) lying on the lower half convex hull of the scatter plot of SSIM and k. For the sake of simplicity, the median point on the convex hull ($k[x^{**}]$, $SSIM[x^{**}]$) is chosen as the new reference view to avoid being too close or too far from the reference view.

\begin{figure}[H]
\centering
\begin{subfigure}{.5\textwidth}
    \centering
    \captionsetup{justification=centering}
    \includegraphics[width=\linewidth]{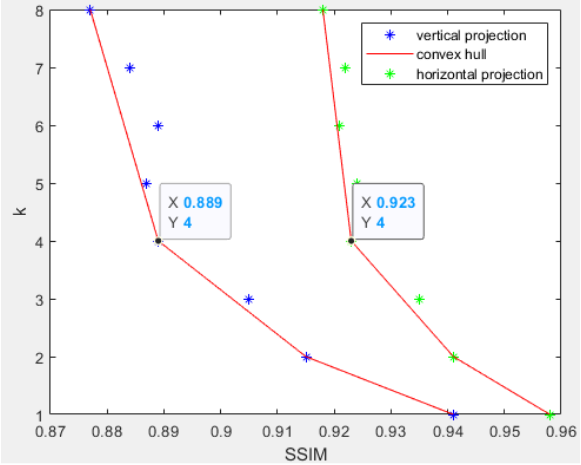}
    \caption{Scatter plot of target views based on its distance to reference view and its projection quality. Potential candidates selected as the next reference view lie on the convex hull \label{fig13a}}
\end{subfigure}%
\begin{subfigure}{.5\textwidth}
    \centering
    \captionsetup{justification=centering}
    \includegraphics[width=.8\linewidth]{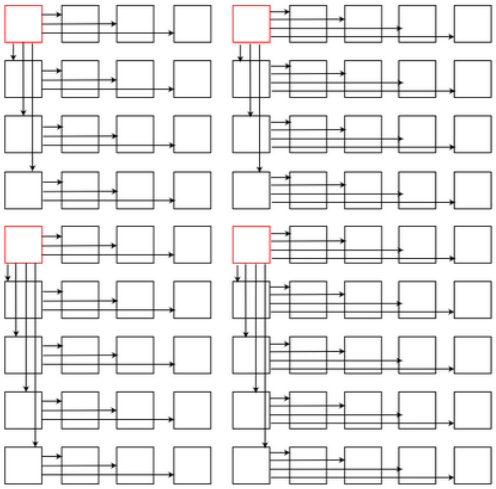}
    \caption{Multi-view projection scheme using every 4th view horizontally and vertically as reference views \label{fig13b}}
\end{subfigure}
\captionsetup{justification=centering}
\caption{Multi-view projection scheme in dataset $Greek$}
\label{fig13}
\end{figure}

An example is shown in Fig. \ref{fig13}, using data in Table \ref{tbl4} of dataset $Greek$ 9x9 views, the SSIM values are plotted against k values for horizontal projection (first row) and vertical projection (first column). For each type of projection, this method finds its corresponding convex hull, then determines the median point as the next reference. Fig. \ref{fig13a} reveals $k[x^{**}]=4$ for both directions, or the new reference for horizontal projection is the 4th view away from the original top-left reference view, while vertical projection also selects the 4th view. New projection scheme can be seen in Fig. \ref{fig13b} with four reference views, instead of one. It should be noted that views $I_{1,9}$ and $I_{5,9}$ are not selected as references because no more views to be projected after them, despite being the 4th view away from the previous references. 

Denote local graph to be the graph with spatial and angular connections within a super-ray, and global graph to be the set of all local graphs. The original high dimensional global graph is now partitioned into four sub global graphs with better projection quality and less complexity, while still exploiting angular correlations of at least four views in every direction. 

Additionally, depending on the technical implementation, all four sub global graphs can be processed simultaneously by taking advantage of parallel programming. The main complexity of graph based LF coding lies in its Laplacian diagonalization of each local graph, which is $O(n^3)$, with $n$ as the number of nodes. By partitioning into four sub global graphs, $n$ is reduced by fourth approximately for each global graph, if not accounting for graph coarsening, and thus time computation can decrease significantly. With graph coarsening enabled to reduce vertices by approximating original graph, as detailed in \cite{ref-15}, the number of nodes in original scheme might be fewer than the total number of nodes in all sub global graphs, because the original graph with higher dimensions may have more coarsened local graphs than each of sub global graphs with lower dimensions in the multi references scheme. However, the number of nodes in each sub global graph is significantly smaller than the original, and each sub global graph is processed independently, hence time computation for both encoder and decoder can still be reduced considerably.

\begin{figure}[H]
\centering
\begin{subfigure}{.5\textwidth}
    \centering
    \captionsetup{justification=centering}
    \includegraphics[width=\linewidth]{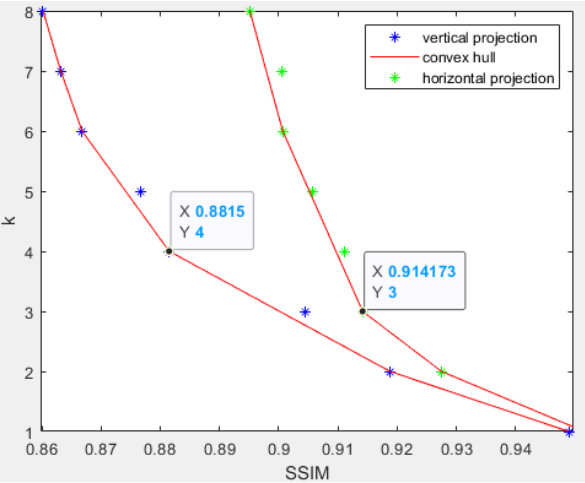}
    \caption{Scatter plot of target views based on its distance to reference view and its projection quality. Potential candidates selected as the next reference view lie on the convex hull \label{fig14a}}
\end{subfigure}%
\begin{subfigure}{.5\textwidth}
    \centering
    \captionsetup{justification=centering}
    \includegraphics[width=.8\linewidth]{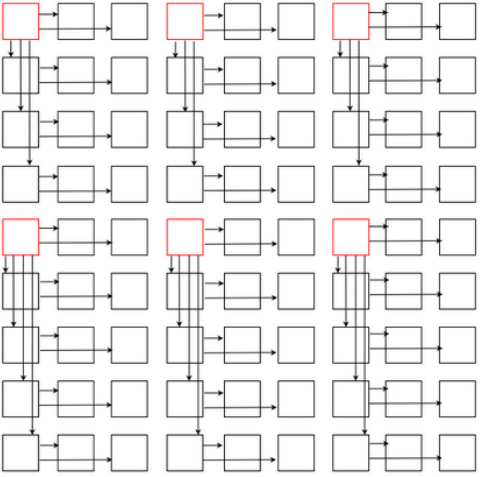}
    \caption{Multi-view projection scheme using every 3rd view horizontally and every 4th view vertically as reference views. \label{fig14b}}
\end{subfigure}
\captionsetup{justification=centering}
\caption{Multi-view projection scheme in dataset $Sideboard$}
\label{fig14}
\end{figure}

The selection of references for dataset $Sideboard$ 9x9 views is shown in Fig. \ref{fig14a} and Fig. \ref{fig14b}, in which horizontal projection selects every 3rd view as the new reference, and vertical projection selects 4th view. The original graph is partitioned into six sub-graphs, exploiting angular correlations of at least three views in every direction, and having better projected segmentation maps for all the views, compared to the original projection scheme. 

\section{Performance Evaluation} \label{results}

In this section, evaluation of super-ray projection quality for each view is analysed quantitatively to show that it can be improved by the two proposed center view, and multiple views projection schemes. Then, a set of experiments are designed to evaluate the impact of enhanced projection quality on overall compression efficiency. Finally, experimental results are presented and analysed.

\subsection{Projection quality evaluation}

\subsubsection{Center view projection scheme}
Using equation (\ref{eq:8}), this experiment computed the projection quality of a vignetted LF, but with center view as the reference. Table \ref{tbl6} and Table \ref{tbl7} show SSIM quality results on all the 13x13 views of $Fountain\_Vincent\_2$ and $Danger\_de\_Mort$, with absolute SSIM on center view. Due to the accurate disparity map, it can be seen that the quality deterioration from the center view to further views horizontally and vertically were slower than in the case of projecting from top-left view, described in Table \ref{tbl2} and Table \ref{tbl3}. Additionally, quality of more views were improved because center view projected to more closer views (smaller disparity) to 4 directions, whereas corner view projected to further views (higher disparity) to 2 directions.

\begin{table}[H]
\centering
\begin{tabular}{*{13}{R}} 
 \hline
0.917 & 0.918 & 0.919 & 0.919 & 0.918 & 0.918 & 0.921 & 0.919 & 0.919 & 0.918 & 0.917 & 0.919 & 0.917 \\ \hline
0.922 & 0.925 & 0.925 & 0.927 & 0.926 & 0.926 & 0.929 & 0.926 & 0.926 & 0.924 & 0.924 & 0.924 & 0.923 \\ \hline
0.927 & 0.928 & 0.929 & 0.929 & 0.929 & 0.931 & 0.930 & 0.928 & 0.927 & 0.926 & 0.925 & 0.926 & 0.927 \\ \hline
0.938 & 0.940 & 0.940 & 0.941 & 0.943 & 0.944 & 0.946 & 0.943 & 0.941 & 0.938 & 0.938 & 0.939 & 0.938 \\ \hline
0.939 & 0.940 & 0.940 & 0.944 & 0.945 & 0.947 & 0.949 & 0.944 & 0.942 & 0.941 & 0.940 & 0.941 & 0.941 \\ \hline
0.955 & 0.954 & 0.955 & 0.959 & 0.962 & 0.968 & 0.979 & 0.965 & 0.960 & 0.959 & 0.957 & 0.955 & 0.954 \\ \hline
0.959 & 0.959 & 0.961 & 0.965 & 0.969 & 0.977 & 1.000 & 0.974 & 0.970 & 0.968 & 0.966 & 0.962 & 0.958 \\ \hline
0.955 & 0.957 & 0.957 & 0.958 & 0.962 & 0.965 & 0.974 & 0.968 & 0.964 & 0.961 & 0.962 & 0.960 & 0.956 \\ \hline
0.943 & 0.944 & 0.945 & 0.946 & 0.946 & 0.948 & 0.953 & 0.949 & 0.947 & 0.947 & 0.944 & 0.943 & 0.940 \\ \hline
0.940 & 0.941 & 0.941 & 0.941 & 0.943 & 0.944 & 0.946 & 0.945 & 0.946 & 0.944 & 0.942 & 0.941 & 0.939 \\ \hline
0.928 & 0.929 & 0.931 & 0.931 & 0.930 & 0.930 & 0.931 & 0.930 & 0.931 & 0.930 & 0.928 & 0.924 & 0.924 \\ \hline
0.922 & 0.925 & 0.927 & 0.928 & 0.928 & 0.927 & 0.927 & 0.926 & 0.925 & 0.924 & 0.920 & 0.921 & 0.918 \\ \hline
0.911 & 0.913 & 0.914 & 0.917 & 0.917 & 0.918 & 0.916 & 0.915 & 0.914 & 0.913 & 0.912 & 0.911 & 0.908 \\ \hline
\end{tabular}
\captionsetup{justification=centering}
\caption{Projection quality of 13x13 views in dataset $Fountain\_Vincent\_2$ using center view projection scheme. Reference view with absolute quality is highlighted in green.}\label{tbl6}
\end{table}

\begin{table}[H]
\centering
\begin{tabular}{ *{13}{R} } 
 \hline
0.917 & 0.916 & 0.917 & 0.920 & 0.917 & 0.916 & 0.919 & 0.920 & 0.918 & 0.917 & 0.916 & 0.918 & 0.916 \\ \hline
0.920 & 0.924 & 0.924 & 0.925 & 0.924 & 0.925 & 0.928 & 0.924 & 0.924 & 0.923 & 0.922 & 0.922 & 0.921 \\ \hline
0.925 & 0.926 & 0.927 & 0.927 & 0.927 & 0.929 & 0.929 & 0.927 & 0.926 & 0.925 & 0.924 & 0.925 & 0.926 \\ \hline
0.936 & 0.939 & 0.939 & 0.939 & 0.943 & 0.942 & 0.944 & 0.941 & 0.939 & 0.937 & 0.937 & 0.938 & 0.936 \\ \hline
0.938 & 0.939 & 0.939 & 0.943 & 0.943 & 0.945 & 0.947 & 0.943 & 0.940 & 0.939 & 0.939 & 0.939 & 0.939 \\ \hline
0.954 & 0.955 & 0.953 & 0.958 & 0.960 & 0.967 & 0.981 & 0.963 & 0.958 & 0.957 & 0.959 & 0.953 & 0.952 \\ \hline
0.957 & 0.957 & 0.960 & 0.963 & 0.968 & 0.976 & 1.000 & 0.973 & 0.968 & 0.966 & 0.965 & 0.961 & 0.956 \\ \hline
0.954 & 0.956 & 0.956 & 0.960 & 0.960 & 0.963 & 0.972 & 0.966 & 0.962 & 0.959 & 0.960 & 0.959 & 0.954 \\ \hline
0.942 & 0.942 & 0.943 & 0.944 & 0.945 & 0.946 & 0.951 & 0.948 & 0.945 & 0.945 & 0.943 & 0.942 & 0.939 \\ \hline
0.938 & 0.939 & 0.941 & 0.939 & 0.942 & 0.942 & 0.945 & 0.943 & 0.944 & 0.943 & 0.941 & 0.940 & 0.937 \\ \hline
0.929 & 0.927 & 0.930 & 0.930 & 0.928 & 0.929 & 0.929 & 0.929 & 0.930 & 0.928 & 0.927 & 0.922 & 0.923 \\ \hline
0.921 & 0.923 & 0.926 & 0.926 & 0.926 & 0.926 & 0.926 & 0.924 & 0.924 & 0.923 & 0.919 & 0.920 & 0.917 \\ \hline
0.909 & 0.912 & 0.912 & 0.916 & 0.915 & 0.917 & 0.914 & 0.913 & 0.913 & 0.912 & 0.910 & 0.910 & 0.906 \\ \hline
\end{tabular}
\captionsetup{justification=centering}
\caption{Projection quality of 13x13 views in dataset $Danger\_de\_Mort$ using center view projection scheme. Reference view with absolute quality is highlighted in green.}\label{tbl7}
\end{table}

\subsubsection{Multiple views projection scheme}

Table \ref{tbl8} and Table \ref{tbl9} show projection quality of synthetic LF $Greek$ and $Sideboard$ using multiple views projection scheme. Absolute SSIM was found in all reference views. Although the deteriorating rate of quality remained fast due to large disparity, quality of remaining views were highly improved, compared to Table \ref{tbl4} and Table \ref{tbl5}. This can be explained by the fact that views with worst projection quality used its ground truth segmentation labels instead, then they became new reference views with accurate segmentation for new projection chains.

\begin{table}[H]
\centering
\begin{tabular}{ *{9}{R} } 
 \hline
1.000 & 0.958 & 0.941 & 0.931 & 1.000 & 0.961 & 0.945 & 0.935 & 0.926 \\ \hline
0.941 & 0.932 & 0.925 & 0.919 & 0.940 & 0.931 & 0.926 & 0.920 & 0.913 \\ \hline
0.915 & 0.912 & 0.908 & 0.906 & 0.913 & 0.912 & 0.910 & 0.905 & 0.900 \\ \hline
0.898 & 0.897 & 0.898 & 0.896 & 0.901 & 0.897 & 0.896 & 0.894 & 0.890 \\ \hline
1.000 & 0.956 & 0.941 & 0.929 & 1.000 & 0.959 & 0.942 & 0.931 & 0.923 \\ \hline
0.933 & 0.925 & 0.920 & 0.917 & 0.936 & 0.927 & 0.922 & 0.917 & 0.912 \\ \hline
0.909 & 0.907 & 0.902 & 0.901 & 0.910 & 0.906 & 0.904 & 0.900 & 0.898 \\ \hline
0.895 & 0.894 & 0.893 & 0.891 & 0.895 & 0.892 & 0.891 & 0.890 & 0.887 \\ \hline
0.888 & 0.885 & 0.882 & 0.883 & 0.887 & 0.887 & 0.885 & 0.883 & 0.880 \\ \hline
\end{tabular}
\captionsetup{justification=centering}
\caption{Projection quality of 9x9 views in dataset $Greek$ using multi-view projection scheme. Reference views with absolute quality are highlighted in green}\label{tbl8}
\end{table}

\begin{table}[H]
\centering
\begin{tabular}{ *{9}{R} } 
 \hline
1.000 & 0.952 & 0.927 & 1.000 & 0.941 & 0.921 & 1.000 & 0.946 & 0.923 \\ \hline
0.949 & 0.934 & 0.921 & 0.942 & 0.925 & 0.912 & 0.946 & 0.928 & 0.911 \\ \hline
0.920 & 0.914 & 0.907 & 0.915 & 0.906 & 0.898 & 0.912 & 0.905 & 0.901 \\ \hline
0.902 & 0.897 & 0.891 & 0.896 & 0.890 & 0.887 & 0.900 & 0.891 & 0.885 \\ \hline
1.000 & 0.945 & 0.925 & 1.000 & 0.945 & 0.918 & 1.000 & 0.943 & 0.915 \\ \hline
0.946 & 0.929 & 0.914 & 0.942 & 0.927 & 0.909 & 0.943 & 0.924 & 0.908 \\ \hline
0.910 & 0.907 & 0.902 & 0.915 & 0.908 & 0.898 & 0.916 & 0.905 & 0.898 \\ \hline
0.894 & 0.889 & 0.889 & 0.897 & 0.893 & 0.889 & 0.898 & 0.891 & 0.887 \\ \hline
0.881 & 0.880 & 0.879 & 0.882 & 0.884 & 0.880 & 0.885 & 0.880 & 0.878 \\ \hline
\end{tabular}
\captionsetup{justification=centering}
\caption{Projection quality of 9x9 views in dataset $Sideboard$ using multi-view projection scheme. Reference views with absolute quality are highlighted in green}\label{tbl9}
\end{table}

\subsection{Compression efficiency evaluation}
In order to evaluate the impact of enhanced projection quality on overall performance, this section assesses Rate Distortion performance and quality of reconstructed LF of the two proposed Center view and Multi-view projection schemes, and time computation for the Multi-view projection scheme. This allows to demonstrate the improvement of quality in both proposals, and also running time for multiple views projection. Real LF datasets captured with plenoptic cameras were downloaded from the EPFL dataset \cite{ref-32}, in which vignetted real LF with large parallax 13x13 views were $Fountain\_Vincent\_2$, and $Danger\_de\_Mort$. Meanwhile, $Greek$ and $Sideboard$ datasets were evaluated for synthetic LF from \cite{ref-33}. In Rate Distortion quantitative results, the proposals were compared against the original top-left view projection scheme and two state-of-the-art coders: HEVC with Serpentine scanning topology, and JPEG Pleno with 4D transform mode (4DTM).

\subsubsection{Experiment Setup}

The encoder and decoder were run on Python 3 under Ubuntu 20.04 with 64GB RAM, and utilized python’s Ray library for the parallel processing of super-rays or sub global graphs. The disparity estimation technique was used from \cite{ref-28} to compute the disparity map for center-view of real LF, and multiple reference views for synthetic LF. Their segmentation mask was obtained using SLIC algorithm \cite{ref-27}. Due to lack of memory resources in the this experimental environment, the initial number of super-rays were set as 2000 and 1200 for real LF 13x13 views and synthetic LF 9x9 views respectively, instead of 500 as originally used in \cite{ref-15}. The more super-rays, the smaller the local graphs, and thus they would consume less resources, but with a trade-off of inefficient decorrelation of signals. On the other hand, having smaller number of super-rays leads to bigger sizes of graphs, which implies significant increase in time complexity of eigen-decomposition for Laplacian matrix.

The subjective results were obtained when running encoder and decoder with parameter PSNRmin set at 20, instead of 45 (max value). This parameter was used to guide the rate of graph coarsening and partitioning. Setting at 20 would return results at low quality (PSNR >= 20), and thus it would be easier to visually differentiate results of the original and proposed projection schemes, for the purpose of reading paper.

The x.265 implementation of HEVC-Serpentine used in the experiments was run with source version 2.3, following LF common test conditions \cite{ref-31}. The base quantization parameters (QP) were set to 10, 28, 35, 50. JPEG Pleno 4DTM was used within Part 4 (Reference Software) with base Lambdas (quantization parameter) of 25, 1000, 20000, 500000.

Regarding QPs used in both original and proposed schemes, this work implemented an adaptive quantization approach. After GFT was used to transform signals into frequency domain, the super-rays coefficients were divided into 32 sub-groups. Since the first group contains low frequency coefficients, which represent fundamental properties of the signals, and it usually has much higher energy than the next groups, hence more quantization  steps should be assigned for the first group to obtain more accurate reconstruction than other groups, for containing more important coefficients. In other words, the base QPs were set adaptively for the first group and remaining groups. Then optimized quantization step sizes were found based on rate-distortion optimization approach, as described in \cite{ref-15}, with parameter QP set according for each group. Using the optimized quantization steps sizes, the coefficients in each group were quantized and arithmetically coded with a public version of Context Adaptive Binary Arithmetic Coder (CABAC) \cite{ref-34}. At high quality coding, the QPs were set as 4 for the first group, and 10 for the remaining groups.
The reference segmentation mask was encoded using arithmetic edge coder EAC \cite{ref-35}, and disparity values (median disparity per super-ray) were encoded using original arithmetic coder of the authors \cite{ref-15}.

Additionally, all reconstructed LFs using any of the four methods in the experiments were converted into 8-bit for evaluation at the same conditions and their PSNR of the luminance channel (PSNR-Y) were computed with the same formula, following the LF common test conditions \cite{ref-31}.

\subsection{Analysis of Center view projection scheme}

\subsubsection{Rate Distortion analysis}
\begin{figure}[H]
\centering
\begin{subfigure}{.5\textwidth}
    \centering
    \captionsetup{justification=centering}
    \includegraphics[width=.9\linewidth]{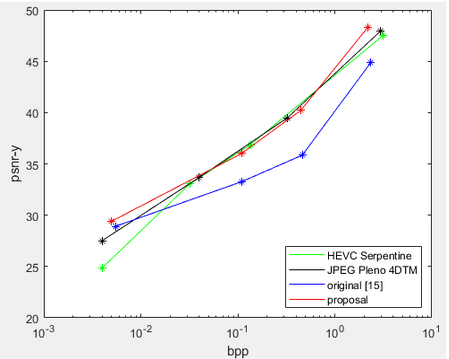}
    \caption{$Fountain\_Vincent\_2$ \label{fig15a}}
\end{subfigure}%
\begin{subfigure}{.5\textwidth}
    \centering
    \captionsetup{justification=centering}
    \includegraphics[width=.9\linewidth]{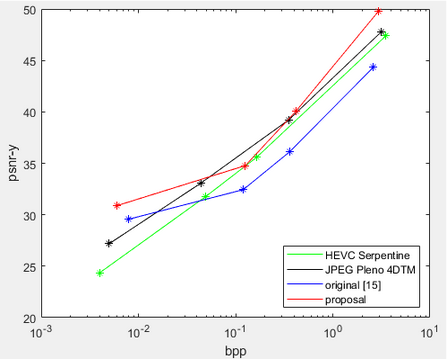}
    \caption{$Danger\_de\_Mort$ \label{fig15b}}
\end{subfigure}
\captionsetup{justification=centering}
\caption{Rate Distortion performance between center view projection scheme (proposal), top-left view projection scheme (original), and state-of-the-art codecs HEVC-Serpentine, JPEG Pleno 4DTM}
\label{fig15}
\end{figure}

Rate Distortion performance of the proposed projection with center view as reference was compared against top-left view projection \cite{ref-13}, direct encoding of the views as a PVS using HEVC-Serpentine, and 4D transform solution utilizing JPEG Pleno 4DTM. The performance comparison was made on the two datasets $Fountain\_Vincent\_2$ and $Danger\_De\_Mort$, as shown in Fig. \ref{fig15}. Substantial gains in the Center view projection proposal can be seen compared to the original scheme at all bitrates for both datasets, and also the proposed scheme outperformed HEVC-Serpentine and JPEG Pleno 4DTM, especially at low and high bitrates. 

\subsubsection{Qualitative analysis for reconstructed LF}
\begin{figure}[H]
\centering
\begin{subfigure}{.5\textwidth}
    \centering
    \captionsetup{justification=centering}
    \includegraphics[width=.95\linewidth]{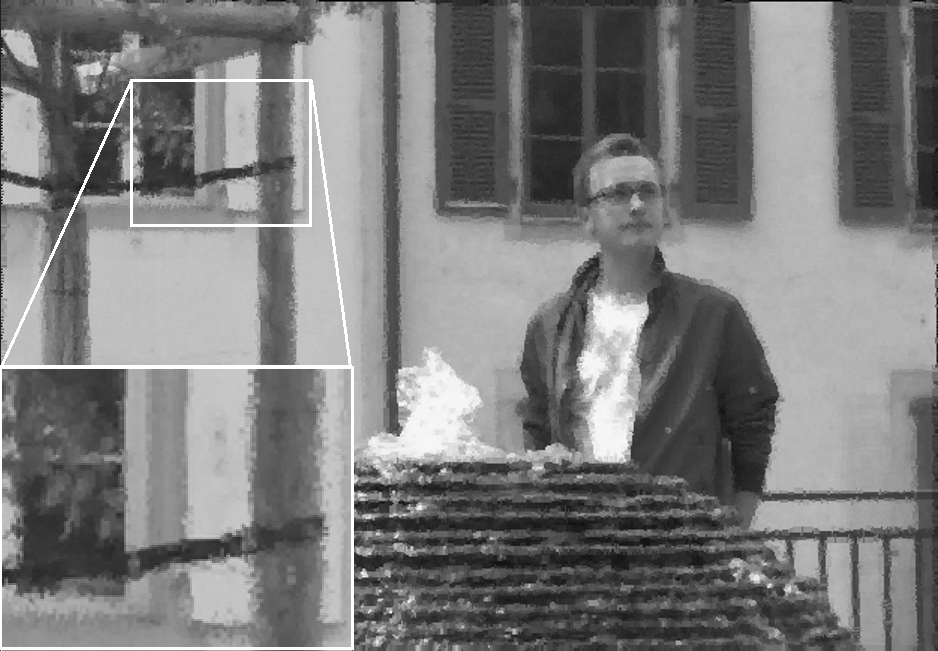}
    \caption{Top-left view projection scheme \cite{ref-15} (original) \label{fig16a}}
\end{subfigure}%
\begin{subfigure}{.5\textwidth}
    \centering
    \captionsetup{justification=centering}
    \includegraphics[width=.95\linewidth]{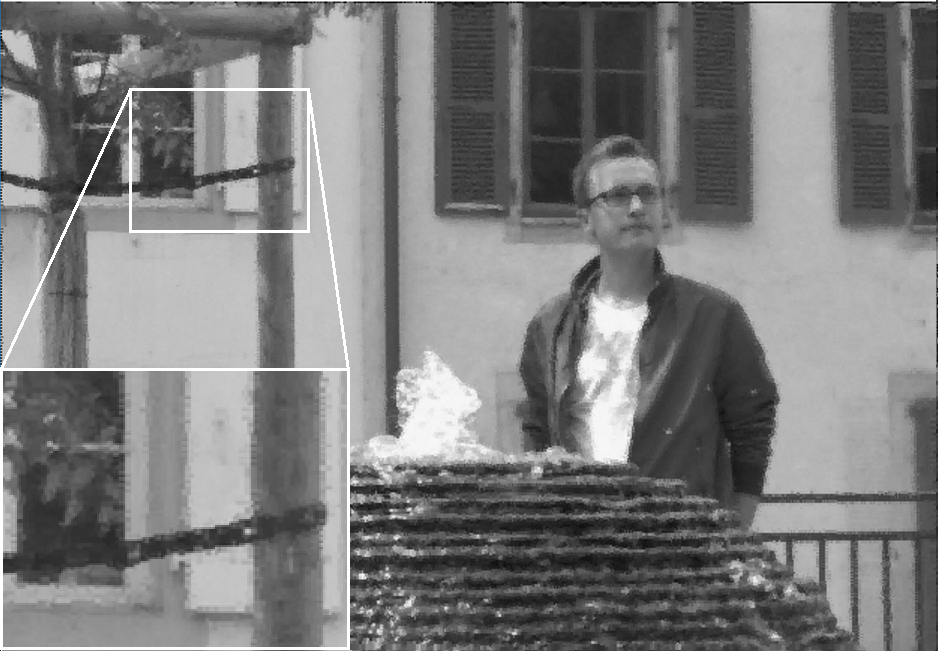}
    \caption{Center view projection scheme (proposal) \label{fig16b}}
\end{subfigure}
\captionsetup{justification=centering}
\caption{Reconstructed luminance channel of center view using projection scheme from top-left view and center view, in dataset $Fountain\_Vincent\_2$}
\label{fig16}
\end{figure}

\begin{figure}[H]
\centering
\begin{subfigure}{.5\textwidth}
    \centering
    \captionsetup{justification=centering}
    \includegraphics[width=.95\linewidth]{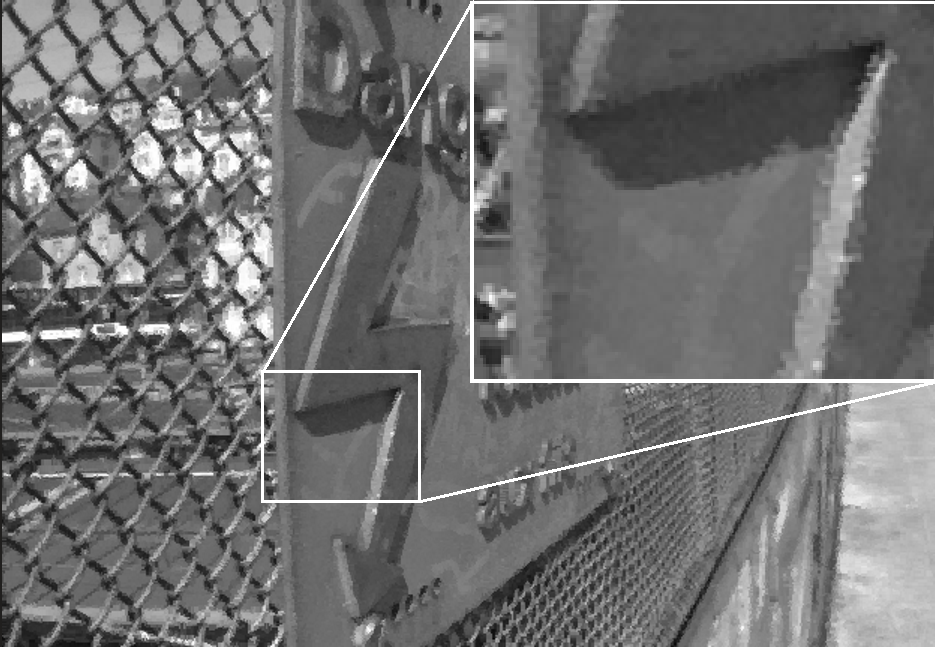}
    \caption{Top-left view projection scheme \cite{ref-15} (original) \label{fig17a}}
\end{subfigure}%
\begin{subfigure}{.5\textwidth}
    \centering
    \captionsetup{justification=centering}
    \includegraphics[width=.95\linewidth]{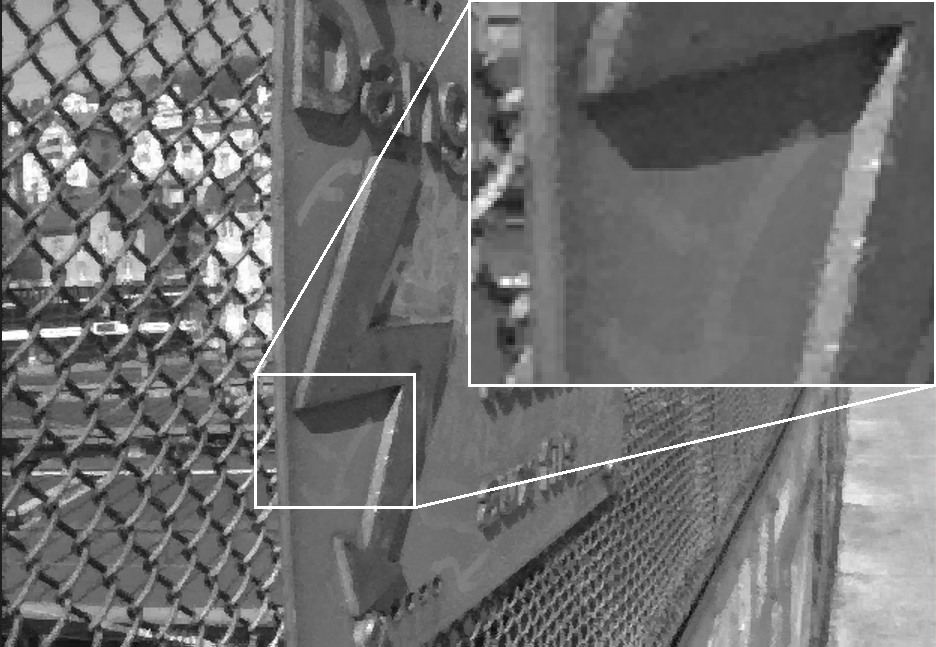}
    \caption{Center view projection scheme (proposal) \label{fig17b}}
\end{subfigure}
\captionsetup{justification=centering}
\caption{Reconstructed luminance channel of center view using projection scheme from top-left view and center view, in dataset $Danger\_de\_Mort$}
\label{fig17}
\end{figure}

At the decoder side, the luminance channel of LF was reconstructed from the quantized coefficients sent by the encoder. The output results using the original and proposed projection scheme are shown subjectively in Fig. \ref{fig16} for $Fountain\_Vincent\_2$, and Fig. \ref{fig17} for $Danger\_de\_Mort$. It can be seen that in both datasets, the proposed Center view projection returned sharper results, clearly visible in edges around texture, whereas the original scheme’s results seemed to be blurry in these edges. The blur effect could be caused by super-pixels reconstructed at inaccurate positions, resulting from poor depth estimation, as a consequence of vignetted top-left view.

\subsection{Analysis of Multiple views projection scheme}

\subsubsection{Rate Distortion analysis}
\begin{figure}[H]
\centering
\begin{subfigure}{.5\textwidth}
    \centering
    \captionsetup{justification=centering}
    \includegraphics[width=.9\linewidth]{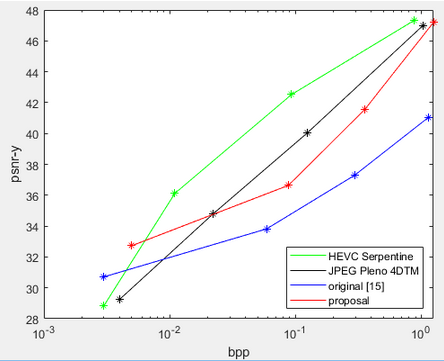}
    \caption{$Greek$ \label{fig18a}}
\end{subfigure}%
\begin{subfigure}{.5\textwidth}
    \centering
    \captionsetup{justification=centering}
    \includegraphics[width=.9\linewidth]{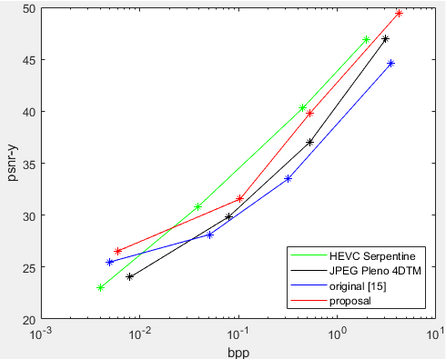}
    \caption{$Sideboard$ \label{fig18b}}
\end{subfigure}
\captionsetup{justification=centering}
\caption{Rate Distortion performance between multi-view projection scheme (proposal), top-left view projection scheme (original), and state-of-the-art codecs HEVC-Serpentine, JPEG Pleno 4DTM}
\label{fig18}
\end{figure}

Rate Distortion performance of the Multi-view projection scheme was illustrated in Fig. \ref{fig18}, comparing with original single view projection scheme, HEVC, and JPEG Pleno in datasets Greek and Sideboard. The proposed scheme significantly outperformed the original projection at all bitrates, for having better projection quality, and surpassed the other two conventional coders at low bitrates. However, HEVC-Serpentine remained the best compressor for synthetic LF at medium and high bitrates. This can be explained by the fact that the two synthetic LF are free of imperfections such as image noises, and thus the performance of classical coders HEVC and JPEG Pleno were not degraded, hence better Rate Distortion than their results in real world LF. Nevertheless, the proposal performed slightly worse on dataset $Greek$, compared to $Sideboard$, because the disparity between views in Greek is higher than other datasets, as shown in Table \ref{tbl1}, leading to higher median disparity error per super-ray used for projection. In addition, more sub-global graphs can be found in $Sideboard$ than $Greeks$ after finding optimized positions for reference views, resulting in more views with better projection quality.

\subsubsection{Qualitative analysis for reconstructed LF}
\begin{figure}[H]
\centering
\begin{subfigure}{.5\textwidth}
    \centering
    \captionsetup{justification=centering}
    \includegraphics[width=.95\linewidth]{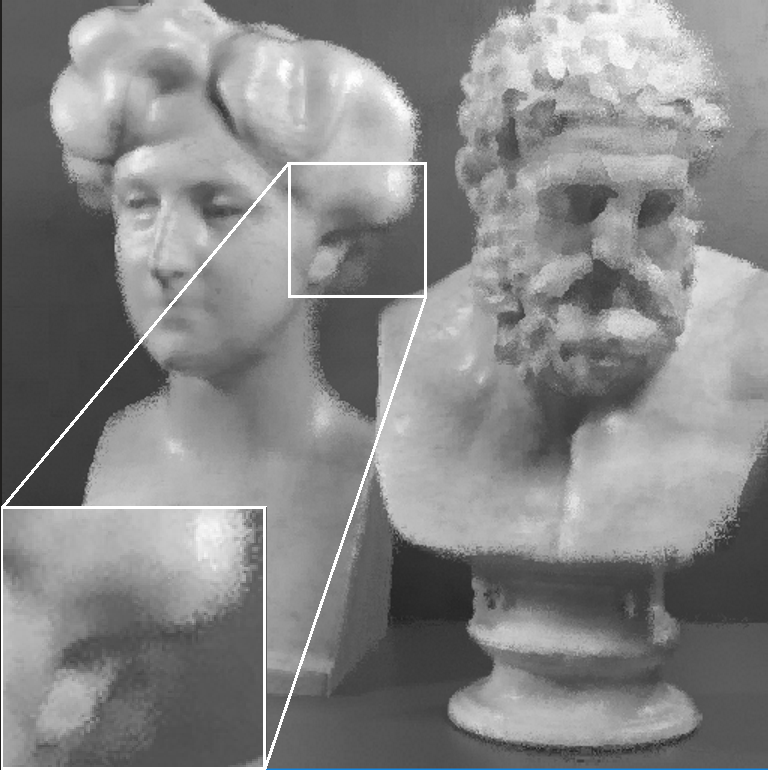}
    \caption{Single view projection scheme \cite{ref-15} (original) \label{fig19a}}
\end{subfigure}%
\begin{subfigure}{.5\textwidth}
    \centering
    \captionsetup{justification=centering}
    \includegraphics[width=.95\linewidth]{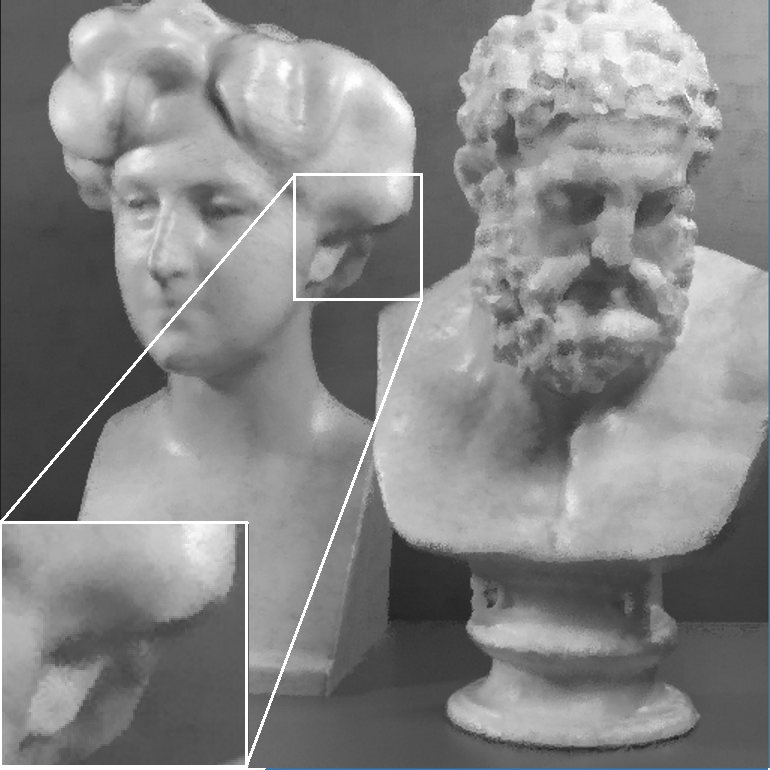}
    \caption{Multiple views references scheme (proposal) \label{fig19b}}
\end{subfigure}
\captionsetup{justification=centering}
\caption{Reconstructed luminance channel of center view using single-view and multi-view projection scheme, in dataset $Greek$}
\label{fig19}
\end{figure}

\begin{figure}[H]
\centering
\begin{subfigure}{.5\textwidth}
    \centering
    \captionsetup{justification=centering}
    \includegraphics[width=.95\linewidth]{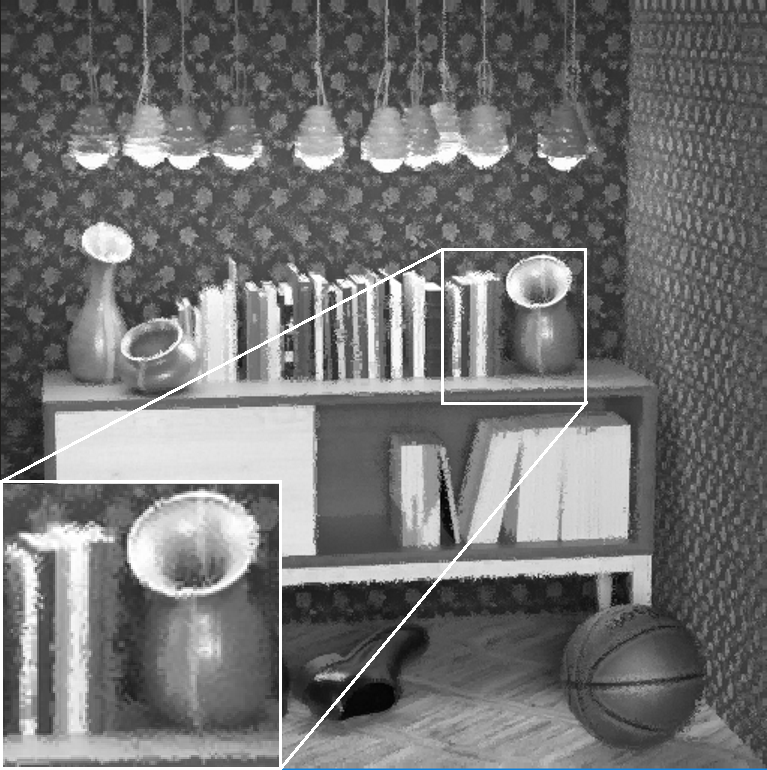}
    \caption{Single view projection scheme \cite{ref-15} (original) \label{fig20a}}
\end{subfigure}%
\begin{subfigure}{.5\textwidth}
    \centering
    \captionsetup{justification=centering}
    \includegraphics[width=.95\linewidth]{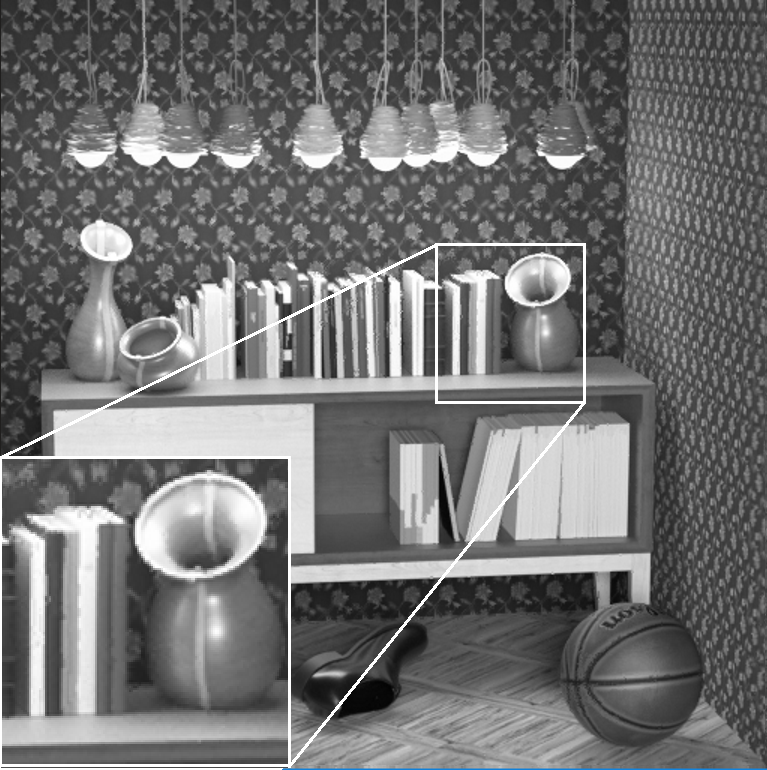}
    \caption{Multiple views references scheme (proposal) \label{fig20b}}
\end{subfigure}
\captionsetup{justification=centering}
\caption{Reconstructed luminance channel of center view using single-view and multi-view projection scheme, in dataset $Sideboard$}
\label{fig20}
\end{figure}

The qualitative results of luminance reconstruction for $Greek$ and $Sideboard$ datasets using original or Multi-view projection schemes are shown in Fig. \ref{fig19} and Fig. \ref{fig20}. Same as previous subjective results of real LF, the proposed Multi-view projection returned sharper results for synthetic LF, especially around edges of textures, for having more accurate projection of super-pixels than the single-view projection scheme.

\subsubsection{Time computation analysis}
Aside from achieving substantial gains in compression performance compared to single-view projection scheme, multi-view projection can also significantly reduce time computation of both encoder and decoder, with a slight trade off increasing bitrates. An example is given in Table \ref{tbl10}, analyzing the parameters and time duration when running encoder and decoder on dataset $Greek$ with PSNRmin set at 40, along with output quality PSNR-Y and required bitrate at high quality coding. The high dimensional graph was separated into four sub global graphs in the multi-view proposal, as optimized by a minimization problem. The first three columns (param, obtained $num\_SR$, total \# of nodes) bring interesting results. It can be seen that, for the same initial $num\_SR$ (number of super-rays / number of local graphs), the output $num\_SR$ and total number of nodes after graph coarsening and partitioning of single-view scheme (original) were much higher than output of each sub global graphs (proposal). This means graph coarsening and partitioning rates were higher in the original graph than in each sub global graph. Thus, the proposed multi-view scheme could retain more accurate graph information of vertice signals and edges in each sub global graph, in addition to having higher quality for the projection of super-pixels, since each reference view projected to closer views. Essentially, the total number of nodes determines the time complexity for eigen-decomposition of the Laplacian matrix, and it was smaller in each sub global graph. Also, the four sub global graphs were encoded or decoded simultaneously by running in parallel, and thus the total approximate encoding and decoding time were reduced by more than half, compared to processing the original high dimensional graph. Nevertheless, the bitrate slightly increased due to the fact that more reference segmentation masks and disparity maps were required to be coded and transmitted alongside the graph coefficients.

\begin{table}[H]
\centering
\captionsetup{justification=centering}
\includegraphics[width=\linewidth]{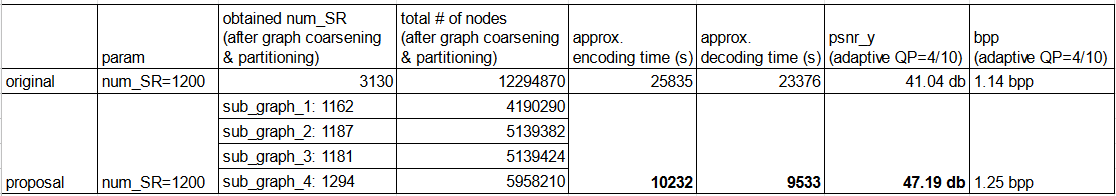}
\caption{Encoding time and Decoding time using single-view (original) and multi-view (proposal) projection scheme on dataset $Greek$ \label{tbl10}}
\end{table}

\section{Discussion} \label{discussion}

Based on evaluation results, it has been shown that the Center view and Multiple views projection scheme can bring an overall improvement for super-rays projection quality in all views by having accurate disparity information, leading to better compression effiency, expecially at high bitrates, compared to the original Top-left view projection. Additionally, combining accurate geometry information with the advantage of graph coarsening proposed in \cite{ref-15}, the graph-based approach can also outperform state-of-the-art coders HEVC and JPEG Pleno at low bitrates. 

The benefit of graph coarsening for graph based approaches is clear, for low bitrates, aside from its ability to exploit correlations for irregular patterns in textures, which allows them to outperform the other two state-of-the-art coders HEVC and JPEG Pleno. Graph coarsening retains total variations of signals on the reduced graphs, while the number of coefficients to be coded also substantially decreases, leading to good Rate Distortion performance at low bitrates. Additionally, beside vignetting effect, real world LF might also suffer from image noises, degrading the performance of traditional coding considerably, but not affecting graph coarsening, which utilizes low rank model approximation, and thus the noises can be removed.

High quality coding at high bitrates requires particularly accurate super-ray positions, which depends entirely on the performance of depth estimation algorithm. For real LF, As verified in previous section, vignetting effect significantly degrades the output depth map of top-left view, leading to inaccurate super-ray projections, hence the original projection scheme obtains the lowest Rate Distortion performance. On the other hand, Center view projection scheme has more accurate depth estimation maps, leading to higher compression performance than HEVC and JPEG Pleno, which might also be supported by its ability to decorrelate signals in irregular-shape textures.
	
For high quality coding of synthetic LF, although the proposed Multiple views projection scheme significantly surpasses the performance of top-left projection scheme, they are still outperformed by HEVC-Serpentine. The potential solution to obtain competitive performance with HEVC is to use more reference views, but with a trade off of increasing bitrates for transmission because more segmentation and disparity information of the references are needed to be coded and sent to the decoder side. Another possible solution could be using two median disparity values for each super-pixel within the reference view, then each half super-pixel would be projected separately to other views, based on the corresponding median disparity value. The idea is motivated by the fact that, the smaller size the super-pixel possesses, the smaller the median disparity error becomes, with respect to all disparity values within the super-pixel. This approach might be discussed further in future work. 
	
Furthermore, for Multi-view projection scheme, time execution for both encoder and decoder can be considerably reduced by processing all sub global graphs in parallel, while ensuring correlations between views can still be efficiently exploited by optimizing positions of reference views through a minimization problem. There may be a slight increase in the coding bitrates due to more reference segmentation and disparity maps are to be coded. Additionally, based on Fig. \ref{fig13a} and \ref{fig14a}, it should be noted that solving the minimization problem to find optimal reference view might not make significant improvement for multiview based LF representation, compared to directly choosing center view of every projection direction as the new reference, since there are only a few views to be evaluated, and most of them lie closely on the convex hull. However, this approach can be applied to lenslet based LF representation, in which the number of views are large, and thus finding the views lying on the convex hull can be more efficient. This idea can also be further discussed in future research.

\section{Conclusion} \label{conclusion}
In this paper, two novel projection schemes for graph-based Light Field coding have been introduced, including Center view and Multiple views projection. The proposals significantly outperformed original Top-left view projection scheme and generally obtained competitive rate-distortion performance with state-of-the-art coders HEVC (Serpentine scanning) and JPEG Pleno (4DTM mode). This can only be achieved by having accurate disparity estimation for Center view projection in real LF with large parallax, and smaller median disparity error for Multiple views projection in synthetic LF. In addition to improving overall compression efficiency, Multiple views projection can also reduce end-to-end time computation by processing smaller sub global graphs in parallel. This has shown the potential of further improvement for graph-based LF coding in order to achieve both competitive performance in both compression efficiency and time computation, compared to state-of-the-art coders.


\authorcontributions{Conceptualization, N.G.B., C.M.T., and T.N.D.; Methodology, N.G.B., C.M.T., and P.X.T.; Supervision, P.X.T. and E.K.; Writing---original draft, N.G.B., C.M.T and T.N.D.; Writing---review and editing, P.X.T. and E.K. All authors have read and agreed to the published version of the manuscript.}

\funding{This research received no external funding.}

\institutionalreview{Not applicable}

\informedconsent{Not applicable}

\dataavailability{Not applicable} 

\conflictsofinterest{The authors declare no conflict of interest.}

\begin{adjustwidth}{-\extralength}{0cm}

\reftitle{References}

\end{adjustwidth}
\end{document}